\documentclass[pra,aps]{revtex4}

\usepackage{amsmath,amssymb}
\usepackage{psfrag}
\usepackage[dvips]{epsfig}
\usepackage{epsfig}

\input psfig.sty

\newcommand{\ba}{\begin{eqnarray}} 
\newcommand{\ea}{\end{eqnarray}} 
\newcommand{\be}{\begin{equation}} 
\newcommand{\ee}{\end{equation}} 
\newcommand{\bea}{\begin{eqnarray}} 
\newcommand{\eea}{\end{eqnarray}} 

\usepackage{theorem}
\newtheorem{theorem}{Theorem}

\newtheorem{definition}{Definition}

\newtheorem{proposition}{Proposition}
\newtheorem{corollary}{Corollary}
\theorembodyfont{\rmfamily}

\theoremstyle{break}
\newtheorem{example}{Example}
\def\QED{~\rule[-1pt]{5pt}{5pt}\par\medskip}

\newenvironment{proof}{{\bf Proof: \ }}{ \hfill \QED}
 
\DeclareMathOperator{\diag}{diag}

\DeclareMathOperator{\tr}{tr}
\DeclareMathOperator{\Ad}{Ad}
\DeclareMathOperator{\ad}{ad}
\DeclareMathOperator{\Ent}{Ent}

\def\R{\mathbb{R}}

\def\C{\mathbb{C}}
\def\Z{\mathbb{Z}}
\def\g{\mathfrak{g}}
\def\p{\mathfrak{p}}
\def\k{\mathfrak{k}}

\def\a{\mathfrak{a}}
\def\su{\mathfrak{su}}
\def\so{\mathfrak{so}}
\def\sl{\mathfrak{sl}}
 
\begin{document} 
\def\CC{{\rm\kern.24em \vrule width.04em height1.46ex 
depth-.07ex \kern-.30em C}}  
\def\RR{{\rm\kern.24em \vrule width.04em height1.46ex 
depth-.07ex \kern-.30em R}}  
\def\II{{\rm\kern.24em \vrule width.04em height1.46ex 
depth-.07ex \kern-.30em 1}}  

\title{A geometric theory of non-local two-qubit operations} 
 
\author{Jun Zhang$^1$, Jiri Vala$^2$, K. Birgitta Whaley$^2$ 
and Shankar Sastry$^1$}
 
\affiliation{$^1$Department of Electrical Engineering 
and Computer Sciences\\ $^2$
Department of Chemistry and Pitzer Center for Theoretical Chemistry\\
University of California, Berkeley, CA 94720}
 
\date{\today}

\begin{abstract} 
We study  non-local two-qubit 
operations from a geometric perspective. 
By applying a Cartan decomposition to 
$\su(4)$, we find that the  
geometric structure of non-local gates is a 3-Torus. 
We derive the invariants for local transformations, and connect these local 
 invariants to the coordinates of the 3-Torus. Since 
 different points on the 3-Torus may correspond to the same local 
 equivalence class, we use the Weyl group theory to reduce the symmetry. 
 We show that the local equivalence classes of two-qubit gates  
are in one-to-one 
 correspondence with the points in a tetrahedron except on the base. 
We then study the properties of perfect 
 entanglers, that is, the two-qubit operations that can  
generate maximally entangled states from some initially separable states. 
We provide criteria to determine whether a given two-qubit gate is a 
perfect entangler and establish 
 a geometric description of perfect entanglers  
by making use of the tetrahedral representation of non-local gates.  
We find that exactly half the non-local gates 
are perfect entanglers. 
We also investigate the non-local operations generated by a 
 given Hamiltonian.  
We first study the gates that can be directly generated by a Hamiltonian.  
Then we explicitly construct a quantum circuit that contains at most three 
non-local gates generated by a two-body interaction Hamiltonian, together 
with at most four local gates generated by single qubit terms.
We prove that such a quantum circuit  can simulate any 
arbitrary two-qubit gate exactly, 
 and hence it provides an efficient implementation of universal 
quantum computation and simulation.
\end{abstract}  
 
\maketitle 
 
\section{Introduction} 
\label{Sec:Introduction} 
Considerable effort has been made on the characterization 
 of non-local properties of quantum states and operations.  
Grassl {\it et al.} \cite{Grassl:98} have computed locally invariant  
polynomial functions of density matrix elements.  
Makhlin \cite{Makhlin:00} has recently analyzed non-local  
properties of two-qubit gates and presented 
local invariants for an operation $M \in U(4)$. 
Makhlin also studied some basic properties of perfect entanglers,  
which are defined as the unitary operations that can generate maximal  
entangled states from some initially separable states. 
Also shown were entangling properties of gates generated by  
several different Hamiltonian operators. 
All these results are crucial for physical implementations  
of quantum computation schemes. 
 
Determining the entangling capabilities of operations generated  
by a given physical system 
is another intriguing and complementary issue. 
Zanardi \cite{Zanardi:00,Zanardi:01} has explored the  
entangling power of quantum evolutions. 
The most extensive recent effort to characterize entangling  
operations is due to Cirac and coworkers \cite{Dur:01a,Cirac:01a, 
Dur:01b,Kraus:01,Vidal:01,Vidal:02a,Vidal:02b,Hammerer:02}.  
Kraus and Cirac \cite{Kraus:01} focused on finding 
the best separable two-qubit input states such that some given unitary  
transformation can create maximal entanglement. 
Vidal {\it et al.} \cite{Vidal:01} developed the interaction cost  
for a non-local operation as the optimal time to generate it  
from a given Hamiltonian. The same group, Hammerer {\it et al.}  
\cite{Hammerer:02} then extended these considerations to  
characterize non-local gates.  
These works are closely related to time optimal control  
as addressed recently by Khaneja {\it et al.} \cite{Khaneja:01}, 
who studied systems described by a Hamiltonian that contains both a 
non-local internal or drift term, and a local control term. 
All these studies assume that any single qubit operation 
can be achieved almost instantaneously.  
This is a good approximation for the situation when the  
control terms in the Hamiltonian can be made large compared  
to the internal couplings. 
 
Universality and controllability are issues of crucial importance in 
 physical implementations of quantum information processing 
 \cite{Gruska:99,Nielsen:00}.  
A series of important results have been obtained since questions 
 of universality were first addressed by Deutsch in his seminal 
 papers on quantum computing \cite{Deutsch:85,Deutsch:89}.
Deutsch \cite{Deutsch:89} 
proved that any unitary operation can be constructed from
generalized Toffoli gate operating on three qubits. 
DiVincenzo \cite{DiVincenzo:95} proved universality for 
 two-qubit gates by reconstructing three-qubit operations using
these gates and a local NOT gate. Similarly,
Barenco \cite{Barenco:95b,Barenco:95} and Sleator and Weinfurter 
 \cite{Sleator:95} identified the controlled unitary operation
as a universal two-qubit gate.  
Barenco \cite{Barenco:95c} showed the universality of 
the CNOT gate supplemented with any single qubit
unitaries, and pointed out advantages of CNOT in the context
of quantum information processing.  
Lloyd \cite{Lloyd:95} showed that almost any quantum 
 gate for two or more qubits is universal. 
Deutsch {\it et al.} \cite{Deutsch:95} proved that almost 
 any two-qubit gate is universal by showing that 
 the set of non-universal operations in $U(4)$ is of 
 lower dimension than the $U(4)$ group.  
Universal properties of quantum gates acting on an $n \ge 2$ 
 dimensional Hilbert space have been studied by Brylinski 
 \cite{Brylinski:01}.  
Dodd {\it et al.} \cite{Dodd:01} have pointed out that 
 universal quantum computation can be achieved by 
 any entangling gate supplemented with local operations. Bremner {\it 
et al.} \cite{Bremner:02} recently demonstrated this by 
extending the results of Brylinski, giving a constructive proof  
that any two-qubit entangling gate can generate CNOT if arbitrary 
single qubit operations are also available. 
Universal sets of quantum gates for $n$-qubit systems have 
 been explored by Vlasov \cite{Vlasov:01,Vlasov:02} in 
 connection with Clifford algebras.  
 
 General results on efficient simulation of any unitary 
 operation in $SU(2^n)$ by a discrete set of 
 gates are embodied in the Solovay-Kitaev theorem 
 \cite{Kitaev:97,Nielsen:00} and in recent work due to 
 Harrow {\it et al.} \cite{Harrow:01}.  
The Solovay-Kitaev theorem implies the equivalence of different 
 designs of universal quantum computers based on suitable discrete  
sets of single-qubit and two-qubit operations in a quantum circuit. 
An example is the standard universal set  
of gates including controlled-NOT and three discrete single-qubit gates,  
namely Hadamard, Phase and $\pi/2$ gates \cite{Nielsen:00}. 
Other universal sets have also been proposed. 
According to the Solovay-Kitaev theorem, 
every such design can represent a circuit that is formulated 
using the standard set of gates. Consequently, all quantum computation 
 constructions---including algorithms, error-correction, and 
 fault-tolerance---can be efficiently simulated 
 by physical systems that can provide a suitable set of 
 operations, and do not necessarily need to be implemented by the
 standard gates.  This  
moves the focus from study of gates to study of the  
Hamiltonians whose time evolution gives rise to the gates.   
In this context, Burkard {\it et al.} \cite{Burkard:99b} 
 studied the quantum computation potential of the 
 isotropic exchange Hamiltonian. This interaction 
 can  generate $\sqrt{\text{SWAP}}$ gate directly.  However, CNOT cannot 
 be obtained directly from the exchange interaction. 
  Burkard {\it et al.} showed that it can be generated via a 
 circuit of two $\sqrt{\text{SWAP}}$ gates and a single-qubit 
 phase rotation.   
More recently, Whaley and co-workers have shown that  
the two-particle exchange interaction is universal when  
physical qubits are encoded into logical qubits, allowing  
a universal gate set to be constructed from this interaction alone 
\cite{Bacon:00,DiVincenzo:00a,Kempe:01a,Kempe:01b,Kempe:01c,Vala:02b}. 
This has given rise to the notion of ``encoded universality'',  
in which a convenient physical interaction is made universal by  
encoding into a subspace \cite{Kempe:01a,Kempe:01b}. 
 Isotropic, anisotropic, and generalized forms of the  
exchange interaction have recently been shown to possess 
 considerable power for efficient construction of universal  
gate sets, allowing explicit universal gate constructions that  
require only a small number of physical operations 
 \cite{ DiVincenzo:00a,Lidar:01c,Kempe:01c,Vala:02b}. 
 
In this paper, we analyze non-local two-qubit operations  
from a geometric perspective and show that considerable insight  
can be achieved with this approach.  
We are concerned with three main questions here. First, achieving a 
geometric representation of two-qubit gates. Second, characterizing or  
quantifying all operations that can generate maximal entanglement.  
Third, exact simulation of any arbitrary 
two-qubit gates from a given two-body physical interaction together with single 
qubit gates. 
The fundamental mathematical techniques we employ are Cartan 
decomposition and Weyl group in the Lie group representation 
theory. The application of these theories to the Lie algebra $\su(4)$ 
provides us with a natural and intuitive geometric approach to investigate the 
properties of non-local two-qubit operations. This geometric approach 
reveals the nature of the problems intrinsically and allows a general formulation of solutions to the three issues of interest here. 
 
Because it is the non-local properties that generate entanglement in quantum 
systems, we first study the invariants and geometric representation of 
non-local two-qubit operations. 
A pair of two-qubit operations are called locally equivalent if they differ 
only by local operations. We apply the Cartan decomposition theorem to 
$\su(4)$, the Lie algebra of the special unitary group $SU(4)$. 
We find that  
the geometric structure of non-local gates is a 3-Torus. 
On the other hand, the Cartan decomposition of $\su(4)$ derived 
from the complexification of $\sl(4)$ yields an easy way to derive 
the invariants for local transformations.  
These invariants can be used to determine whether two 
gates are locally equivalent. Moreover, we establish the relation of these 
local invariants to the coordinates of the 3-Torus. This provides with a 
way to compute the corresponding points on the 3-Torus for a given 
gate. It turns out that a single non-local gate may correspond to 
finitely many different points on the 3-Torus. If we represent these 
points in a cube with side length $\pi$, there is obvious 
symmetry between these points. 
We then use the Weyl group theory to reduce this symmetry. We know that 
in this case the Weyl group is generated by a set of  
reflections in $\R^3$. It is these 
reflections that create the kaleidoscopic symmetry of points that 
correspond to the same non-local gate in the cube. 
We can explicitly compute these reflections, and thereby  
show that the local equivalence classes of two-qubit gates are in one-to-one 
correspondence with the points in a tetrahedron except on the base. 
This provides a complete geometric representation of non-local two-qubit 
operations.  
 
The second objective of this paper is to explore the 
properties of perfect  
entanglers, that is, the quantum gates that can  
generate maximally entangled states from some initially separable 
states. We start with criteria to  
determine whether a given two-qubit gate is a 
perfect entangler. A condition for such a gate has been stated in  
~\cite{Makhlin:00}.  We provide here a proof of this condition and show that 
the condition can be employed within our geometric analysis to  
determine which fraction of all non-local two-qubit gates are  
perfect entanglers.  We show that the entangling property of a 
quantum gate is only determined by its geometric representation on the 
3-Torus.  Using the result that every point on the tetrahedron 
corresponds to a local equivalence class, we then  
show that the set of all perfect entanglers is a polyhedron with 
seven faces and possessing a volume equal to exactly half that of the 
tetrahedron. This implies that amongst all the non-local two-qubit 
operations, exactly half of them are capable of generating maximal 
entanglement. 
 
Finally, we explore universality and controllability aspects of non-local  
properties of given physical interactions and the potential of  
such specified Hamiltonians to generate perfect entanglers.  Our 
motivation is related to that of encoded universality, namely, 
 determining the potential for universal quantum computation and 
simulation of a given physical Hamiltonian.  However, whereas  
encoded universality sought to construct encodings to achieve  
universality of quantum logic, here we focus on the simulation  
of any arbitrary two-qubit gate.  Achievement of this, together  
with our second result above, allows generation of maximal 
entanglement as well as providing universality.  To realize this, we consider here the  
conventional scenario of a Hamiltonian acting on a physical set of qubits   
such that any arbitrary single qubit operation and certain specific  
two-qubit operations may be turned on for selected time durations in series.  
Generally speaking, two-qubit interactions include both local and non-local  
terms. The non-local terms can give rise to not only well-known entangling gates such 
as CNOT, but also to many other classes of gates that may or may  
not lie in the perfect entangling sector.  We therefore seek a 
systematic way to construct quantum circuits from a 
 given physical Hamiltonian that can simulate  
{\em any} arbitrary two-qubit gate exactly, without restriction by  
limitations of speed due to time-scales of control interactions  
(e.g. as in NMR) or to fundamental considerations~\cite{Lloyd:00}.  
As in the study of encoded universality we start with the gates that  
can be directly generated by a given 
Hamiltonian. Generally, these 
gates form a one dimensional subset on the 3-Torus.  
To construct an exact simulation of any arbitrary two-qubit gate, we  
make use of the quantum circuit model. We explicitly construct  
a quantum circuit that contains three non-local gates 
generated by a given two-body interaction Hamiltonian for  
corresponding finite time durations, together with at most  
four local gates. We prove that such a quantum circuit can  
simulate {\em any} arbitrary two-qubit operation exactly and is
therefore universal.  
 In particular, it can therefore {\em efficiently} provide maximal 
 entanglement from any arbitrary Hamiltonian of this form. 
  Such efficient construction from any given Hamiltonian is  
extremely useful for design and experimental implementation  
of quantum information processing schemes.  

\section{Preliminaries}
\label{Sec:Preliminaries}
In this section, we briefly review some basic facts about Cartan
decomposition and the Weyl group within Lie group representation theory
~\cite{Helgason:78, Cahn:84,Hsiang:98}, and then apply these results to
$\su(4)$, the Lie algebra of the special unitary group $SU(4)$.
Applications of Cartan decomposition to quantum system control can also be
found in~\cite{Khaneja:01}.

We concentrate on $SU(4)$ when studying two-qubit gates. 
It is well-known that 
an arbitrary two-qubit gate $U_0\in U(4)$
can be decomposed as the product of a gate $U_1\in SU(4)$ and a global
phase shift $e^{i\alpha}$, where $\alpha\in \R$.
Because the global phase has no significance 
in quantum mechanics, we can thereby reduce the study of the group
$U(4)$ of two-qubit quantum evolution operators to $SU(4)$.
Extensions of results from the group $SU(4)$ back to $U(4)$
are  made when appropriate.

\begin{figure}[t] 
\begin{center} 
 \footnotesize 
 \psfrag{A}[][]{$\text{Local Gates}$} 
 \psfrag{B}[][]{$\text{Non-Perfect Entanglers}$} 
 \psfrag{C}[][]{$\text{Perfect Entanglers}$} 
 \psfrag{E}[][]{$\text{CNOT}$} 
 \psfrag{I}[][]{$\text{Non-local Gates}$} 
 \psfrag{H}[][]{$SU(4)\backslash SU(2)\otimes SU(2)$} 
 \psfrag{G}[][]{$SU(2)\otimes SU(2)$} 
  \psfrag{D}[][]{$\text{SWAP}$} 
\includegraphics[width=0.35\hsize]{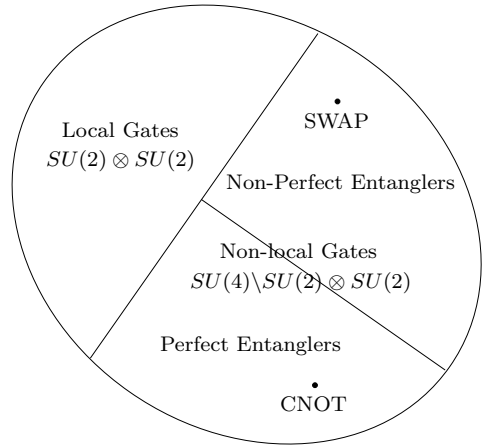}  
\end{center} 
\caption{Partition of all the gates in $SU(4)$} 
 \label{fig:partition} 
\end{figure}

We heuristically introduce a partition of 
the set of two-qubit operations represented by the group $SU(4)$.
This set splits into two subsets, one of 
local gates $SU(2)\otimes SU(2)$ and the other of non-local gates 
$SU(4)\backslash SU(2)\otimes SU(2)$. The latter splits further into
a set of perfect entanglers, {\it i.e.}, those that can generate maximally entangled 
states, an example of which is CNOT, 
and the complementary set of those non-local gates that are not perfect entanglers.
This schematic partition is illustrated in Fig. \ref{fig:partition}.
A rigorous definition of perfect entanglers is presented in Sec. \ref{Sec:Perfect}. 

\subsection{Cartan decomposition and Weyl group}
\label{Sec:Cartan}
Our first goal is to establish fundamentals for a geometric picture of 
non-local unitary operations with emphasis on their generators, which 
 are represented by Hamiltonian operators in physical context.
We start with a summary of  some basic definitions~\cite{Helgason:78, Cahn:84,Hsiang:98}. 
Consider a Lie group $G$ and its corresponding 
Lie algebra $\g$. The adjoint representation 
$\Ad_g$ is a map from the Lie algebra $\g$ to $\g$ which is the
differential of the conjugation map
$a_g$ from the Lie group $G$ to $G$ given by $a_g(h) =  g h g^{-1}$.
For matrix Lie algebras, $\Ad_g (Y) = g Y g^{-1}$,
 where $g$, $Y$ are both represented as matrices of
compatible dimensions. The differential of the adjoint representation
is denoted by $\ad$, and $\ad_X$ is a map from the Lie algebra $\g$
to $\g$ given by the Lie bracket with $X$, that is,
$ad_X (Y) = [X, Y]$.

We now define an inner product on $\g$ 
by the Killing form $B(X, Y)=\tr(\ad_X \ad_Y)$.
Let $\{X_1, \dots, X_n\}$ be a basis for $\g$. The 
numbers $C_{jk}^i\in \C$ such that 
\begin{equation}
  \label{eq:44}
    [X_j, X_k]=\sum_{i=1}^n C_{jk}^i X_i
\end{equation}
are the \emph {structure constants} of the Lie algebra $\g$ with respect to
the basis, where $j$, $k$ run from $1$ to $n$. 
Since 
\begin{eqnarray}
  \label{eq:80}
\aligned
  \ad_{X_j}[X_1, \dots, X_n]&=[\sum_{i=1}^n C_{j1}^i X_i, \dots,
\sum_{i=1}^n C_{jn}^i X_i]\\
&=[X_1, \dots, X_n]\left[
    \begin{matrix}
      C_{j1}^1&\cdots& C_{jn}^1\\
      \vdots&&\vdots \\
      C_{j1}^n&\cdots& C_{jn}^n
    \end{matrix}
\right],
\endaligned
\end{eqnarray}
the matrix representation of $\ad_{X_j}$ with respect to the basis is 
\begin{equation}
  \label{eq:52}
\left[    \begin{matrix}
      C_{j1}^1&\cdots& C_{jn}^1\\
      \vdots&&\vdots \\
      C_{j1}^n&\cdots& C_{jn}^n
    \end{matrix}
\right].
\end{equation}
Thus, the trace of $\ad_{X_j}\ad_{X_k}$, which is $B(X_j, X_k)$, is
$\sum_{a, b=1}^n\ C_{jb}^aC_{ka}^b$, which is also the $jk$-th entry
of the matrix of the quadratic form $B(\ , \ )$.
The Lie algebra $\g$ is semisimple if and only if the Killing form is
nondegenerate, {\it i.e.}, the determinant of its matrix is nonzero.
  
Let $K$ be a compact subgroup of $G$, and $\k$ the Lie algebra 
of $K$. Assume that $\g$ admits a direct sum decomposition 
$\g=\p\oplus\k$, such that $\p=\k^\bot$ with
respect to the metric induced by the inner product.
\begin{definition}[Cartan decomposition of the Lie algebra $\g$]
Let $\g$ be a semisimple Lie algebra and let the decomposition
$\g=\p\oplus \k$, $\p=\k^\bot$ satisfy the commutation relations  
\begin{equation}
  \label{eq:42}
  [\k, \k]\subset \k,\quad [\p, \k]\subset\p,\quad [\p, \p]\subset \k.
\end{equation}
This decomposition is called a Cartan decomposition of
$\g$, and the pair $(\g, \k)$ is called an orthogonal symmetric Lie
algebra pair.
\end{definition}
A maximal Abelian subalgebra $\a$ contained in $\p$ is 
called a \emph{Cartan subalgebra} of the pair $(\g,\k)$. 
If $\a'$ is another Cartan
subalgebra of $(\g,\k)$, then there exists an element $k\in K$ such that
  $\Ad_k(\a)=\a'$. Moreover, we have $\p=\cup_{k\in K}\Ad_k(\a)$.
  
\begin{proposition}[Decomposition of the Lie group $G$]
\label{prop:cartan}
  Given a semisimple Lie algebra $\g$ and its Cartan decomposition
  $\g=\p\oplus\k$, let $\a$ be a Cartan subalgebra of the pair $(\g,
  \k)$, then $G=K\exp(\a)K$.
\end{proposition}

For $X\in \a$, let $W\in \g$ be an 
eigenvector of $\ad_X$ and $\alpha (X)$ 
the corresponding eigenvalue, {\it i.e.},
\begin{equation}
  \label{eq:34}
[X, W]=\alpha (X)W.  
\end{equation}
The linear function $\alpha$ is called a \emph{root} of $\g$ with
 respect to $\a$.
 Let $\Delta$ denote the set of nonzero roots,
and $\Delta_{\p}$ denote the set of roots in $\Delta$ which do not
vanish identically on $\a$.
Note that if $\alpha\in \Delta$, it is also true that $-\alpha\in \Delta$. 

Let $M$ and $M'$ denote the centralizer and normalizer of $\a$ in
$K$, respectively. In other words,
\begin{eqnarray}
  \label{eq:33}
  \aligned
M&=\{k\in K | \Ad_k(X)=X \text{ for each } X\in \a\},\\
M'&=\{k\in K | \Ad_k(\a) \subset \a\}.
\endaligned
\end{eqnarray}
\begin{definition}[Weyl group]
  The quotient group $M'/M$ is called the \emph{Weyl group} of the pair
  $(G, K)$. It is denoted by $W(G, K)$.
\end{definition}
One can prove that $W(G, K)$ is a finite group. 
Each $\alpha\in \Delta_{\p}$ defines a hyperplane $\alpha(X)=0$ in the
vector space $\a$. These hyperplanes divide the space $\a$ into
finitely many connected components, called the \emph{Weyl chambers}.
For each $\alpha\in \Delta_{\p}$, 
let $s_\alpha$ denote the reflection
with respect to the hyperplane $\alpha(X)=0$ in $\a$.
\begin{proposition}[Generation of the Weyl group]
  \label{prop:weyl}
The Weyl group is generated by the reflections $s_\alpha$, $\alpha\in
\Delta_{\p}$.
\end{proposition}
This proposition is proved in Corollary 2.13, Ch. VII in \cite{Helgason:78}.

\subsection{Application to $\su(4)$}
\label{Sec:CartanSu4}
Now we apply the above results to $\su(4)$, the Lie algebra of the
special unitary group $SU(4)$.
The Lie algebra $\g=\su(4)$ has a direct sum decomposition
$\g=\p\oplus\k$, where
\begin{equation}
  \label{eq:29}
\gathered
 \k = \text{span } \frac{i}{2}\{\sigma_x^1,\ \sigma_y^1,\  \sigma_z^1,\
  \sigma_x^2,\  \sigma_y^2,\  \sigma_z^2\}, \\
\p = \text{span } \frac{i}{2}\{\sigma_x^1\sigma_x^2,\  \sigma_x^1\sigma_y^2,\
\sigma_x^1\sigma_z^2,\  \sigma_y^1\sigma_x^2,\  \sigma_y^1\sigma_y^2,\
\sigma_y^1\sigma_z^2,\  \sigma_z^1\sigma_x^2,\ \sigma_z^1\sigma_y^2,\ 
\sigma_z^1\sigma_z^2 \}.
\endgathered
\end{equation}
Here $\sigma_x$, $\sigma_y$, and $\sigma_z$
 are the Pauli matrices, and
$\sigma_{\alpha}^1\sigma_{\beta}^2 = \sigma_{\alpha}^1 \otimes 
\sigma_{\beta}^2$. 
If we use $X_j$ to denote the matrices in Eq. (\ref{eq:29}), where $j$
runs from left to right in Eq. (\ref{eq:29}),
we can derive the Lie brackets of $X_j$ and $X_k$. These are
 summarized in the following table:
\vspace{0.2cm}
\begin{eqnarray*}
\begin{array}{c|cccccc|ccccccccc}
[X_j, X_k]&X_1&X_2&X_3&X_4&X_5&X_6&X_7&X_8&X_9&X_{10}&X_{11}
 &X_{12}&X_{13}&X_{14}&X_{15}\\ \hline
 X_1&0&-X_3&X_2&0&0&0&0&0&0&-X_{13}&-X_{14}&-X_{15}&X_{10}&X_{11}&X_{12}\\
 X_2&X_3&0&-X_1&0&0&0&X_{13}&X_{14}&X_{15}&0&0&0&-X_7&-X_8&-X_9\\
 X_3&-X_2&X_1&0&0&0&0&-X_{10}&-X_{11}&-X_{12}&X_7&X_8&X_9&0&0&0\\
 X_4&0&0&0&0&-X_6&X_5&0&-X_9&X_8&0&-X_{12}&X_{11}&0&-X_{15}&X_{14}\\
 X_5&0&0&0&X_6&0&-X_4&X_9&0&-X_7&X_{12}&0&-X_{10}&X_{15}&0&-X_{13}\\
 X_6&0&0&0&-X_5&X_4&0&-X_8&X_7&0&-X_{11}&X_{10}&0&-X_{14}&X_{13}&0\\ \hline
 X_7&0&-X_{13}&X_{10}&0&-X_9&X_8&0&-X_6&X_5&-X_3&0&0&X_2&0&0\\
 X_8&0&-X_{14}&X_{11}&X_9&0&-X_7&X_6&0&-X_4&0&-X_3&0&0&X_2&0\\
 X_9&0&-X_{15}&X_{12}&-X_8&X_7&0&-X_5&X_4&0&0&0&-X_3&0&0&X_2\\
 X_{10}&X_{13}&0&-X_7&0&-X_{12}&X_{11}&X_3&0&0&0&-X_6&X_5&-X_1&0&0\\
 X_{11}&X_{14}&0&-X_8&X_{12}&0&-X_{10}&0&X_3&0&X_6&0&-X_4&0&-X_1&0\\
 X_{12}&X_{15}&0&-X_9&-X_{11}&X_{10}&0&0&0&X_3&-X_5&X_4&0&0&0&-X_1\\
 X_{13}&-X_{10}&X_7&0&0&-X_{15}&X_{14}&-X_2&0&0&X_1&0&0&0&-X_6&X_5\\
 X_{14}&-X_{11}&X_8&0&X_{15}&0&-X_{13}&0&-X_2&0&0&X_1&0&X_6&0&-X_4\\
 X_{15}&-X_{12}&X_9&0&-X_{14}&X_{13}&0&0&0&-X_2&0&0&X_1&-X_5&X_4&0
\end{array}
\end{eqnarray*}
\vspace{0.2cm}
Now the structure constants $C_{jk}^i$ can be found from the above
table (see Eq. (\ref{eq:44})) so that we can evaluate
\begin{equation}
  \label{eq:82}
  B(X_j, X_k)=\sum_{a=1}^{15}\sum_{b=1}^{15} C_{jb}^aC_{ka}^b=-8\delta_{jk}.
\end{equation}
It is easy to verify that $\tr(X_jX_k)=-\delta_{jk}$, and thus
the Killing form of $\su(4)$ is $B(X, Y)=8\tr(XY)$. 
Since $\k=\text{span}\,\{X_1, \dots, X_6\}$ and 
$\p=\text{span}\,\{X_7, \dots, X_{15}\}$, from the Lie bracket
computation table above, it is clear that
 \begin{equation}
   \label{eq:58}
   [\k, \k]\subset \k,\quad [\p, \k]\subset \p,\quad [\p, \p]\subset \k.
\end{equation}
Therefore the decomposition $\g=\k\oplus\p$ is a Cartan decomposition
of $\su(4)$. Note that the Abelian subalgebra 
\begin{equation}
  \label{eq:59}
  \a = \text{span } \frac{i}{2}\{\sigma_x^1\sigma_x^2,\ 
\sigma_y^1\sigma_y^2,\sigma_z^1\sigma_z^2\}
\end{equation}
is contained in $\p$ and is a maximal Abelian subalgebra,
{\it i.e.}, we cannot find any other Abelian subalgebra of $\p$ that contains $\a$.
Hence it is a Cartan subalgebra
 of the pair $(\g,\k)$. Further, 
since the set of all the local gates $K$ is a
connected Lie subgroup $SU(2)\otimes SU(2)$ 
of $SU(4)$, and there is one-to-one
correspondence between connected Lie subgroups of a Lie group and
subalgebras of its Lie algebra~\cite{Warner:83}, it is clear that $\k$
in Eq. (\ref{eq:29}) is just the Lie subalgebra corresponding to
$K$. From
Proposition~\ref{prop:cartan}, any $U\in SU(4)$ can be decomposed as 
\begin{equation}
  \label{eq:1}
  U=k_1Ak_2=k_1\exp\{\frac{i}{2}(c_1\sigma_x^1\sigma_x^2 
+c_2\sigma_y^1\sigma_y^2
+c_3\sigma_z^1\sigma_z^2)\} k_2,
\end{equation}
where $k_1$, $k_2\in {SU(2)\otimes SU(2)}$, and $c_1$, $c_2$, $c_3\in \R$.

Another more intuitive Cartan decomposition of $\su(4)$ can be
obtained via the complexification of $\sl(4)$. Consider $G=SL(4)$,
the real special linear group, and $K=SO(4)$, the special orthogonal
group. The Lie algebra $\sl(4)$ is the set of
$4\times 4$ real matrices of trace zero, and $\so(4)$ the set of
$4\times 4$ real skew symmetric matrices. Then $\sl(4)$ can be
decomposed as $\sl(4)=\so(4)\oplus \p$, where $\p$ is the set of
$4\times 4$ real symmetric matrices. This is nothing but the
decomposition of a matrix into symmetric and skew symmetric parts, and
it is indeed a Cartan decomposition of $\sl(4)$. Consider the 
following subset of the complexification of $\sl(4)$:
\begin{equation}
  \label{eq:20}
  \g_\mu=\so(4)+i\p.
\end{equation}
It can be verified that $\g_\mu$ is exactly $\su(4)$, 
and thus $(\g_\mu, \so(4))$
is an orthogonal symmetric Lie algebra pair. The isomorphism carrying
$\k$ in Eq. (\ref{eq:29}) into $\so(4)$ is just the transformation from the
standard computational basis of states to the Bell basis
in~\cite{Makhlin:00, Hammerer:02}. This procedure is of crucial
importance in computing the invariants for two-qubit gates under local
transformations. See Section~\ref{sec:non-local} for more details.

Now let us compute the Weyl group $W(G, K)$. Let 
$X=\frac{i}{2}(c_1\sigma_x^1\sigma_x^2 
+c_2\sigma_y^1\sigma_y^2+c_3\sigma_z^1\sigma_z^2)\in \a$.
Identify $\a$ with $\R^3$, then $X=[c_1, c_2, c_3]$.
The roots of $\g$ with respect to $\a$ are eigenvalues of the matrix
of $\ad_X$:
\begin{eqnarray}
  \label{eq:36}
\aligned
\Delta_{\p}&= i\{c_1-c_2, -c_1-c_2, -c_1-c_3, c_1-c_3, c_2-c_3, c_2+c_3,\\
&\quad -c_1+c_2, c_1+c_2,  c_1+c_3, -c_1+c_3, -c_2+c_3,  -c_2-c_3\}.
\endaligned
\end{eqnarray}
For $\alpha=i(c_1-c_3)\in \Delta_{\p}$, the plane $\alpha(X)=0$ in $\a$
 is the set $\{X\in \R^3| u^T X=0\}$, where $u=[1, 0, -1]^T$.
The reflection of $X=[c_1, c_2, c_3]$
with respect to  the plane $\alpha(X)=0$ is
\begin{equation}
  \label{eq:35}
  s_{\alpha}(X)=X-\frac{2uu^T}{\|u\|^2}X
=[c_3, c_2, c_1].
\end{equation}
Similarly, we can compute all the reflections $s_\alpha$ as follows:
\begin{eqnarray}
  \label{eq:37}
\gathered
s_{i(c_3-c_2)}(X)=[c_1, c_3, c_2],\quad  s_{i(c_2+c_3)}(X)=[c_1, -c_3, -c_2],\\
s_{i(c_2-c_1)}(X)=[c_2, c_1, c_3],\quad  s_{i(c_1+c_2)}(X)=[-c_2,-c_1, c_3],\\
s_{i(c_1-c_3)}(X)=[c_3, c_2, c_1],\quad  s_{i(c_1+c_3)}(X)=[-c_3, c_2, -c_1].
\endgathered
\end{eqnarray}
From Proposition~\ref{prop:weyl}, the Weyl group $W(G, K)$ is
generated by $s_\alpha$ given in Eq. (\ref{eq:37}).
Therefore, the reflections $s_\alpha$ are equivalent to either
permutations of the elements of $[c_1, c_2, c_3]$, or
permutations with sign flips of two elements.

\section{Non-local operations}
\label{sec:non-local}
We  now study  non-local two-qubit
operations within the group theoretical framework of the previous section. 
The Cartan decomposition of $\su(4)$ provides us with a good
starting point to explore the invariants under local gate
operations. It also reveals that the geometric structure of the local
equivalence classes is none other than a 3-Torus. Every point on this
3-Torus corresponds to a local equivalence class of two-qubit gates.
Different points may also correspond to the same equivalence
class. To reduce this symmetry, we apply the Weyl group theory. We
show that the local equivalence classes of two-qubit gates are in one-to-one
 correspondence with the points in a tetrahedron, except on the
 base where there are two equivalent areas.
 This tetrahedral representation of non-local operations plays a
 central role in our subsequent discussion of perfect entanglers and the design
 of universal quantum circuits.

\subsection{Local invariants and local equivalence classes}
Two unitary transformations $U$, $U_1\in SU(4)$ are called \emph{locally
  equivalent} if they differ only by local operations:
$U= k_1 U_1 k_2$,
  where $k_1$, $k_2\in SU(2)\otimes SU(2)$ are local 
  gates. This clearly defines an equivalence relation on
  the Lie group $SU(4)$. We denote the equivalence class of a
  unitary transformation $U$ as $[U]$.
From the Cartan decomposition of $\su(4)$ in
 Section~\ref{Sec:CartanSu4}, any two-qubit gate $U\in
SU(4)$ can be written in the following form
\begin{equation}
  \label{eq:12}
    U=k_1Ak_2=k_1\exp\{\frac{i}{2}(c_1\sigma_x^1\sigma_x^2 
+c_2\sigma_y^1\sigma_y^2+c_3\sigma_z^1\sigma_z^2)\} k_2,
\end{equation}
where $k_1$, $k_2\in {SU(2)\otimes SU(2)}$.
Because the two-qubit gate $U$ is periodic in
$c_k$ and the minimum positive period is $\pi$,
the geometric structure of $[c_1, c_2, c_3]$ is 
a 3-Torus, $T^3=S^1\times S^1\times S^1$. 

In \cite{Makhlin:00}, local invariants were given for 
two-qubit gates.  
Here we will connect these invariants of Makhlin to
the coordinates $[c_1, c_2, c_3]$ on the 3-Torus.
We first consider the case of the two-qubit gates in $SU(4)$, 
and then extend the results to the general case of  $U(4)$.

\subsubsection{SU(4) Operations}
Consider the transformation 
from the standard basis of states $|00\rangle$, 
$|01\rangle$, $|10\rangle$, $|11\rangle$ to the Bell basis
$|\Phi^+\rangle = \frac1{\sqrt{2}}(|00\rangle+|11\rangle)$,
$|\Phi^-\rangle = \frac{i}{\sqrt{2}}(|01\rangle+|10\rangle)$,
$|\Psi^+\rangle = \frac1{\sqrt{2}}(|01\rangle-|10\rangle)$,
$|\Psi^-\rangle = \frac{i}{\sqrt{2}}(|00\rangle-|11\rangle)$. 
In this basis, the two-qubit gate $U$ in Eq. (\ref{eq:12})
can be written as
\begin{equation}
  \label{eq:61}
    U_B=Q^\dag UQ=Q^\dag k_1Ak_2Q,
\end{equation}
where
\begin{equation}
  \label{eq:4}
    Q=\frac1{\sqrt{2}}\left(
  \begin{matrix}
    1&0&0&i\\
0&i&1&0\\
0&i&-1&0\\
1&0&0&-i
  \end{matrix}\right).
\end{equation}
Recalling that $ \frac{i}{2}\{\sigma_x^1, \sigma_y^1,  \sigma_z^1,
  \sigma_x^2, \sigma_y^2, \sigma_z^2\}$ is a basis for
  $\k$, it is not hard to  verify that 
$ \frac{i}{2}Q^\dag\{\sigma_x^1, \sigma_y^1,  \sigma_z^1,
  \sigma_x^2, \sigma_y^2, \sigma_z^2\}Q$
forms a basis for $\so(4)$, the Lie algebra of the special orthogonal
 group $SO(4)$. Hence $U_B$ can be written as 
 \begin{equation}
   \label{eq:57}
   U_B=O_1Q^\dag AQO_2,
\end{equation}
where 
\begin{eqnarray}
  \label{eq:60}
\gathered 
 O_1=Q^\dag k_1Q\in SO(4),\\
O_2=Q^\dag k_2Q\in SO(4).
\endgathered
\end{eqnarray}
Eq. (\ref{eq:57}) can also be obtained from the 
Cartan decomposition of $\su(4)$ derived
from the complexification of $\sl(4)$, as discussed in Section
~\ref{Sec:CartanSu4}. An Abelian subalgebra $\a$
is generated by $\frac{i}2\{\sigma_x^1\sigma_x^2, \sigma_y^1\sigma_y^2,
\sigma_z^1\sigma_z^2\}$, and the transformation to the Bell basis
takes these operators to
  $\frac{i}2\{\sigma_z^1, -\sigma_z^2,\sigma_z^1\sigma_z^2\}$. 
Therefore, we have $U_B=O_1FO_2$,
where
\begin{equation}
    \label{eq:5}
\aligned
    F&=Q^\dag AQ\\
&=\exp\{\frac{i}2(c_1\sigma_z^1-c_2\sigma_z^2+c_3\sigma_z^1\sigma_z^2)\}\\
&=\diag\{e^{i\frac{c_1-c_2+c_3}2}, e^{i\frac{c_1+c_2-c_3}2},
e^{-i\frac{c_1+c_2+c_3}2}, e^{i\frac{-c_1+c_2+c_3}2} \}.
\endaligned
  \end{equation}
Let 
\begin{equation}
  \label{eq:83}
m=U_B^TU_B=O_2^TF^2O_2,  
\end{equation}
where $O_2$ is defined by Eq.~(\ref{eq:60}).
The complete set of local
invariants of a two-qubit gate $U\in SU(4)$ is given by the
spectrum of the matrix $m$~\cite{Makhlin:00}, and hence by the eigenvalues of $F^2$:
\begin{equation}
  \label{eq:62}
  \{\,e^{i(c_1-c_2+c_3)},\ e^{i(c_1+c_2-c_3)},\ 
e^{-i(c_1+c_2+c_3)},\ e^{i(-c_1+c_2+c_3)} \}.
\end{equation}
Since $m$ is unitary and $\det{m}=1$, the characteristic polynomial of
$m$ is then 
\begin{equation}
  \label{eq:30}
|sI-m|=s^4-\tr(m)\, s^3+\frac12\big(\tr^2(m)-\tr(m^2)\big)s^2
-\overline{\tr(m)}\ s+1.
\end{equation}
Therefore the spectrum of $m$ is completely determined by 
only the two quantities $\tr(m)$ and $\tr^2(m)-\tr{(m^2)}$. For a two-qubit gate $U$
given in Eq. (\ref{eq:12}), its local invariants can be 
derived from Eq. (\ref{eq:62}) as:
\begin{eqnarray}
\aligned
\label{eq:9}
\tr{(m)}&=4\cos c_1\cos c_2\cos c_3+4i\sin c_1\sin c_2\sin c_3,\\
\tr^2{(m)}-\tr{(m^2)}&=16\cos^2 c_1\cos^2 c_2\cos^2 c_3-16\sin^2 c_1\sin^2
c_2\sin^2 c_3-4\cos 2c_1\cos 2c_2\cos 2c_3.
\endaligned
\end{eqnarray}

\subsubsection{Generalization to U(4)}
Now let us consider the local invariants for the general case of
$U(4)$ \cite{Makhlin:00}.  
An arbitrary two-qubit gate $U\in U(4)$
can be decomposed as the product of a gate $U_1\in SU(4)$ and a global
phase shift $e^{i\alpha}$, where $\det U=e^{i4\alpha}$. 
It follows that $m({U_1})=e^{-i2\alpha} m({U})$, 
where 
\begin{equation}
  \label{eq:88}
m(U)=(Q^\dag UQ)^T Q^\dag UQ,  
\end{equation}
and 
\begin{eqnarray}
  \label{eq:63}
  \aligned
  \tr\big(m({U_1})\big)&= e^{-i2\alpha}\tr\big( m({U})\big),\\
\tr^2\big( m(U_1)\big)-\tr{\big(m^2(U_1)\big)}&=e^{-i4\alpha}\big(\tr^2(m(U))
-\tr({m^2(U)})\big).
\endaligned
\end{eqnarray}
It is clear that the global phase factor just rotates 
the eigenvalues of $m(U)$ along the unit circle in the complex
plane, while keeping their relative phase invariant.
Therefore, it does not affect the entangling properties and
we can consequently divide by $\det(U)$. 
The local invariants of a two-qubit gate $U$ are thus given by 
\begin{eqnarray}
  \label{eq:32}
\aligned
  G_1&=\frac{\tr^2\big(m(U)\big)}{16\det U},\\
G_2&=\frac{\tr^2\big(m(U)\big)-\tr\big(m^2(U)\big)}{4\det U},
\endaligned
\end{eqnarray}
where the numerical factors are incorporated into the denominators 
to provide convenient normalization.
If $U$ is now written in the following form:
\begin{equation}
  \label{eq:31}
     U=e^{i\alpha}k_1Ak_2=e^{i\alpha}
k_1\exp\{\frac{i}{2}(c_1\sigma_x^1\sigma_x^2 
+c_2\sigma_y^1\sigma_y^2+c_3\sigma_z^1\sigma_z^2)\} k_2,
\end{equation}
we can compute its local invariants as:
\begin{eqnarray}
  \label{eq:11}
  \aligned
G_1&=\cos^2 c_1\cos^2 c_2\cos^2 c_3-\sin^2 c_1\sin^2 c_2\sin^2 c_3
+\frac{i}4 \sin 2c_1\sin 2c_2\sin 2c_3\\
G_2&=4\cos^2 c_1\cos^2 c_2\cos^2 c_3-4\sin^2 c_1\sin^2
c_2\sin^2 c_3-\cos 2c_1\cos 2c_2\cos 2c_3.
\endaligned
\end{eqnarray}

Because the local invariants $G_1$ and $G_2$
characterize the non-local properties of unitary operations,
we can use these two invariants to check whether
a pair of two-qubit gates are locally equivalent.
The invariants $G_1$ and $G_2$ are evaluated by taking the matrix representation
of a gate in the Bell basis and then using Eqs.~(\ref{eq:88}) and (\ref{eq:32}).
For example, CNOT and Controlled-Z (referred as C(Z)) 
possess identical values of the local invariants, given by 
$G_1 = 0$ and $G_2 = 1$. Therefore, they belong to the 
same local equivalence class. We refer to this class as [CNOT].
 On the other hand, the local invariants for $\sqrt{\text{SWAP}}$ 
are $G_1 = i/4$ and $G_2 = 0$. Hence this gate
belongs to a different local equivalence class 
that we refer to as $[\sqrt{\text{SWAP}}]$.
Note that from Eq. (\ref{eq:32}), 
since the local invariants are functions of eigenvalues
 of the matrix $m$, the local equivalence class
can alternatively be defined simply via the set of eigenvalues of the matrix $m$.

\subsection{Geometric representation of two-qubit gates}
\label{sec:geo}
Eq. (\ref{eq:11}) reveals the relation between the local
invariants $G_1$ and $G_2$ 
and the coordinates $[c_1, c_2, c_3]$ of the 3-Torus
structure of non-local two-qubit gates. From this relation, 
given a set of coordinates $[c_1, c_2, c_3]$, we can
easily compute the local invariants for a local equivalence
class. Vice versa, from a given pair of values of the local invariants $G_1$ and $G_2$, 
we can also find the points on the 3-Torus that correspond to a given
two-qubit operation.
In general, we expect to find multiple
points on the 3-Torus for a given pair $G_1$ and $G_2$.  We now show how this
multiple-valued nature can be removed by using the Weyl group to construct
a geometric representation that allows the symmetry to be reduced.

To visualize the geometric structure of the two-qubit gates, we first
consider a cube with side length $\pi$ in the vector
space $\a$. This provides an equivalent representation of the points on the
3-Torus, since $T^3\cong \R^3/\Z^3$. Clearly, every point in this
cube corresponds a local equivalence class. However, different points
in the cube may belong to the same local equivalence class. For
example, both the points $[\frac{\pi}4, \frac{\pi}4, \frac{\pi}4]$
and $[\frac{\pi}4, \frac{3\pi}4, \frac{3\pi}4]$ correspond to the gate
 $\sqrt{\text{SWAP}}$.

We use the theory of the Weyl group to reduce
this symmetry in the cube.
From the Lie group representation theory, the orbits of local
gates $K$ acting on $SU(4)/SU(2)\otimes SU(2)$ are in one-to-one
correspondence with the orbits of the Weyl group $W(G, K)$ on
$\a$~\cite{Helgason:78}. 
From Proposition~\ref{prop:weyl}, the Weyl group $W(G, K)$ is
generated by the reflections $s_\alpha$ as given in Eq. (\ref{eq:37}).
Note that in Eq. (\ref{eq:37}), the reflections $s_\alpha$ are either
permutations or  permutations with sign flips of
two entries in $[c_1, c_2, c_3]$. Therefore, if $[c_1, c_2, c_3]$ 
is an element in a local equivalence
class $[U]$, then $[c_i, c_j, c_k]$, $[\pi-c_i, \pi-c_j, c_k]$, 
$[c_i, \pi-c_j, \pi-c_k]$, and $[\pi-c_i, c_j, \pi-c_k]$ are also in $[U]$, where
$(i, j, k)$ is a permutation of $(1, 2, 3)$. 
With the meaning clear
from the context of the discussion, in the remainder of this paper
we shall 
use the triplet $[c_1, c_2, c_3]$ to denote either the
corresponding local equivalence class of a two-qubit gate, or simply to refer to
a specific point on the 3-Torus or cube.

\begin{figure}[tb]
\begin{center}
\begin{tabular}{ccc}
 \scriptsize
 \psfrag{c1}[][]{$c_1$}
  \psfrag{c2}[][]{$c_2$}
 \psfrag{c3}[][]{$c_3$}
 \psfrag{pi1}[][]{$\pi$}
 \psfrag{pi2}[][]{$\pi$}
 \psfrag{pi3}[][]{$\pi$}
   \psfrag{A1}[][]{$A_1$}
  \psfrag{O}[][]{$O$}
\includegraphics[width=0.4\hsize]{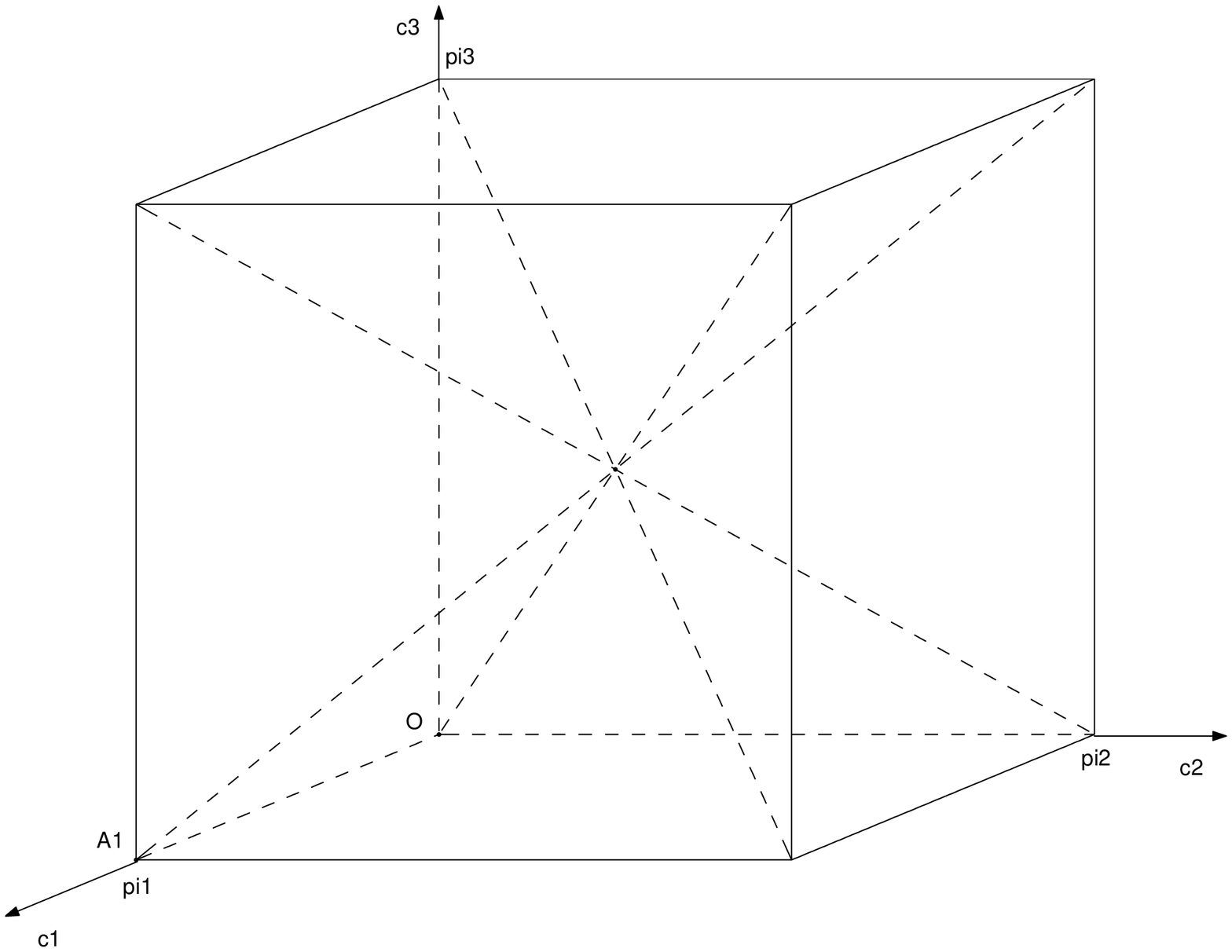} 
&\quad 
\scriptsize
   \psfrag{A1}[][]{$A_1$}
 \psfrag{A2}[][]{$A_2$}
 \psfrag{A3}[][]{$A_3$}
 \psfrag{O}[][]{$O$}
\includegraphics[width=0.3\hsize]{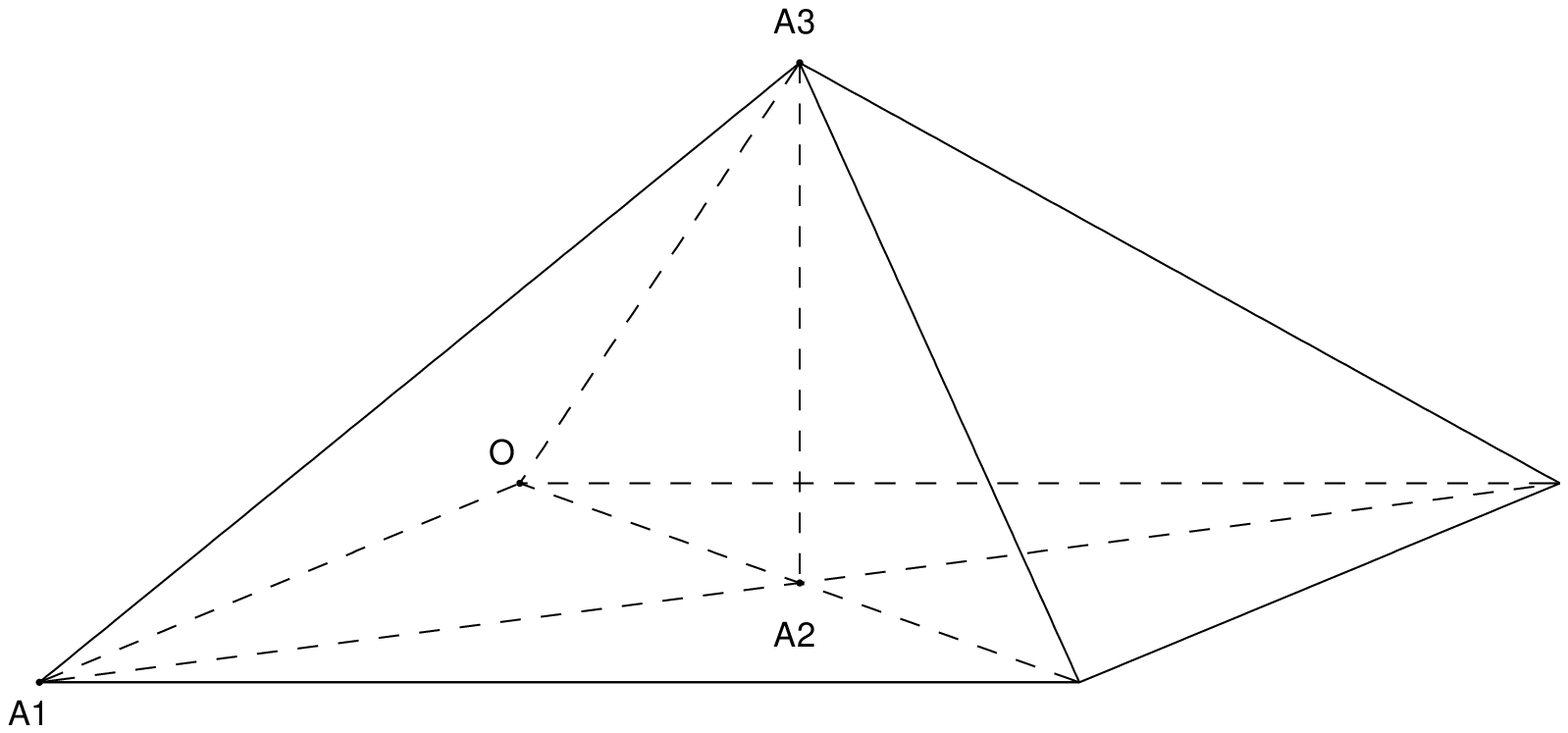} 
&\quad
\scriptsize
   \psfrag{A1}[][]{$A_1$}
 \psfrag{A2}[][]{$A_2$}
 \psfrag{A3}[][]{$A_3$}
 \psfrag{O}[][]{$O$}
 \psfrag{L}[][]{$L$}
\includegraphics[width=0.15\hsize]{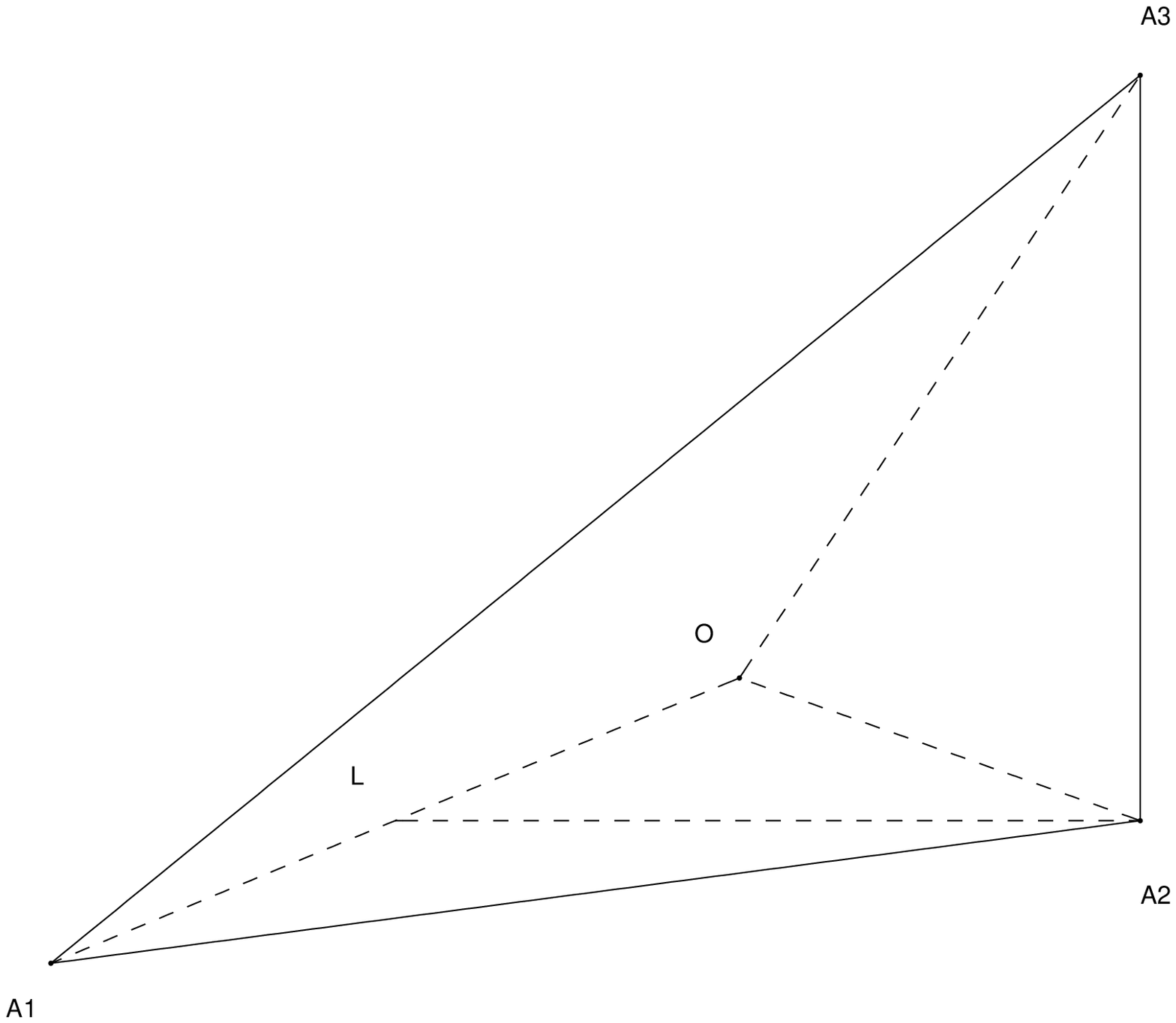} \\
(A)&(B)&(C)
\end{tabular}
\end{center}
 \caption{
Illustration of the tetrahedral representation of non-local
two-qubit operations.
(A) Divide the cube by the planes  $c_1-c_3=0$, $c_1+c_3=\pi$, 
$c_2-c_3=0$, and $c_2+c_3=\pi$.
(B) One of the six equivalent square pyramids produced from (A).  Further
dividing this pyramid by the planes $c_1-c_2=0$ and $c_1+c_2=\pi$ gives
(C), the tetrahedron $\text{OA}_1\text{A}_2\text{A}_3$, 
with $\text{A}_1=[\pi, 0, 0]$, $\text{A}_2=[\frac{\pi}2,
 \frac{\pi}2, 0]$, and $\text{A}_3=[\frac{\pi}2, \frac{\pi}2,
 \frac{\pi}2]$. $\text{OA}_1\text{A}_2\text{A}_3$
 is a Weyl chamber, denoted $a^+$,  
with the exception of points on its base where
we have an equivalence of $\text{LA}_2\text{A}_1$ with
$\text{LA}_2\text{O}$, where L is the point
$[\frac{\pi}2, 0, 0]$.   Every point in $a^+$
corresponds to a local equivalence class of two-qubit operations.
}
 \label{fig:2}
\end{figure}

Since each orbit of the Weyl group $W(G, K)$ on $\a$
contains precisely one point in a Weyl chamber, the
local equivalence classes of two-qubit gates are in one-to-one
correspondence with the points of a Weyl chamber.
Hence, each Weyl chamber contains all the local equivalence classes.
Recall that the Weyl chambers are obtained 
by dividing the vector space $\a$ by
the hyperplanes $\alpha(X)=0$, where $\alpha\in \Delta_{\p}$ as given in
Eq. (\ref{eq:36}). Therefore, we can obtain the Weyl chambers by dividing
the cube by the planes
\begin{eqnarray}
  \label{eq:38}
\gathered
\{X\in \a: c_1-c_2=0\},\quad \{X\in \a: c_1+c_2=\pi\},\\
\{X\in \a: c_1-c_3=0\},\quad \{X\in \a: c_1+c_3=\pi\},\\
\{X\in \a: c_2-c_3=0\},\quad \{X\in \a: c_2+c_3=\pi\}.
\endgathered
\end{eqnarray}
Figure~\ref{fig:2}(A) shows that 
after dividing the cube by the planes $c_1-c_3=0$, $c_1+c_3=\pi$, 
$c_2-c_3=0$, and $c_2+c_3=\pi$, we obtain six square pyramids.
One of these pyramids is shown in Figure~\ref{fig:2}(B). Further
dividing this pyramid by the planes
$c_1-c_2=0$ and $c_1+c_2=\pi$, we get a tetrahedron
$\text{OA}_1\text{A}_2\text{A}_3$ such as that shown in Figure~\ref{fig:2}(C). 
Notice that for any point $[c_1, c_2, 0]$ on the base of this
tetrahedron, its mirror image with respect
to the line $\text{LA}_2$, which is $[\pi-c_1, c_2, 0]$, corresponds
to the same local equivalence class. Therefore, with the caveat that
the basal areas $\text{LA}_2\text{A}_1$ and $\text{LA}_2\text{O}$ 
are identified as equivalent, we finally arrive at the
identification of the tetrahedron $\text{OA}_1\text{A}_2\text{A}_3$
as a Weyl chamber, and we denote this $\a^+$. 
There are $24$ such Weyl chambers in total, and each of them has the volume 
$\pi^3/24$. Note that every point in $\a^+$ corresponds to a different local
equivalence class.  Consequently, the Weyl chamber $\a^+$ provides
a geometric representation of all the possible two-qubit gates. 

For a given two-qubit gate, it is important to find its
coordinates  $[c_1, c_2, c_3]$ on the 3-Torus and hence 
in the Weyl chamber $\a^+$. 
With this representation in the tetrahedron $\a^+$ we 
have removed the multiple-valued nature of the coordinates 
on the 3-Torus and cube
and therefore can now take the coordinates $[c_1, c_2, c_3]$ 
as an alternative set
of local invariants.  They provide a useful geometric representation of local
invariants that is easy to visualize and is entirely equivalent to 
$G_1$ and $G_2$.
They can be used directly to implement the local equivalence class of
particular prescribed two-qubit gates for a given Hamiltonian.
More generally, this alternative set of local invariants helps us to
gain a better understanding of the local invariants and geometric
representation of two-qubit gates. 

It is clear that the local gates $K$
correspond to the points $\text{O}$ and $\text{A}_1$ in
Figures~\ref{fig:2} (A)--(C). 
We now study several nontrivial examples 
of non-local gates to
determine the corresponding coordinates $[c_1, c_2, c_3]$
in $\a^+$. All the other  points of a particular 
local equivalence class
in the cube can be obtained by applying the Weyl group $W(G, K)$
to the corresponding point in $\a^+$. 
Note that for a gate $[c_1, c_2, c_3]$ in $\a^+$, its inverse is just
$[\frac{\pi}2-c_1, c_2, c_3]$.

\noindent{\bf{(1) CNOT}}

Following the procedure to compute the local invariants 
described above (see Eqs. (\ref{eq:88}) and (\ref{eq:32})), 
we obtain $G_1=0$ and $G_2=1$ for the two-qubit gate CNOT. 
Solving Eq. (\ref{eq:11}):
\begin{eqnarray}
  \label{eq:3}
\gathered
\cos^2 c_1\cos^2 c_2\cos^2 c_3-\sin^2 c_1\sin^2 c_2\sin^2 c_3=0,\\
\sin 2c_1\sin 2c_2\sin 2c_3=0,\\
-\cos 2c_1\cos 2c_2\cos 2c_3=1,
\endgathered
\end{eqnarray}
we find that $[\frac\pi{2}, 0, 0]$ is the corresponding point for 
CNOT in the Weyl chamber $\a^+$. This is
 the point $\text{L}$ in Figure~\ref{fig:2}(C).

\noindent{\bf{(2) SWAP}}

For the gate SWAP, we have $G_1=-1$ and $G_2=-3$.
Solving Eq. (\ref{eq:11}), 
we obtain that  the corresponding point for SWAP
is $[\frac{\pi}2, \frac\pi{2}, \frac\pi{2}]$, {\it i.e.}, the point $\text{A}_3$ 
in Figure~\ref{fig:2}(C).

\noindent{\bf{(3) $\sqrt{\text{SWAP}}$}}

The local invariants for the gate $\sqrt{\text{SWAP}}$ are
$G_1=\frac{i}4$ and $G_2=0$. Solving Eq. (\ref{eq:11}) for
this case, we derive that $[\frac{\pi}4, \frac{\pi}{4}, \frac{\pi}{4}]$
is the corresponding point in $\a^+$. This is the midpoint of
$\text{OA}_3$ in Figure \ref{fig:2}(C). 

\noindent{\bf{(4) Controlled-$U$}}

Suppose $U$ is an arbitrary single qubit unitary operation:
\begin{equation}
  \label{eq:64}
    U=\exp(\gamma_1i\sigma_x+\gamma_2i\sigma_y+\gamma_3i\sigma_z).
\end{equation}
For the Controlled-$U$ gate, the local invariants are
$G_1=\cos^2\gamma$ and $G_2=2\cos^2\gamma+1$, where
$\gamma=\sqrt{\gamma_1^2 +\gamma_2^2+\gamma_3^2}$. By solving 
 Eq. (\ref{eq:11}), we find that $[\gamma, 0 , 0 ]$ is the
 corresponding point in $\a^+$. Hence, all the Controlled-$U$
 gates correspond to the line $\text{OL}$ in $\a^+$, 
where $\text{L}$ is  [CNOT].

\section{Characterization of perfect entanglers}
\label{Sec:Perfect}
Entanglement is one of the most striking quantum mechanical features
that plays a key role in quantum computation and
quantum information. It is used in many applications
such as teleportation and quantum cryptography~\cite{Nielsen:00}. In
many applications, it is often desired to generate maximal
entanglement from some unentangled initial states.
The non-local two-qubit operations that can generate maximal
entanglement are called perfect entanglers. 
In this section, we study the perfect entanglers using the geometric
approach established in the preceding sections.
We will prove a theorem that provides a sufficient and necessary
condition for a two-qubit gate to be a perfect entangler.
It turns out that whether a two-qubit gate can generate maximal
entanglement is only determined by its location on the 3-Torus,
or more specifically, in the Weyl chamber $\a^+$. We show that 
in the tetrahedral representation of non-local gates summarized in
Figure~\ref{fig:2}(C), all the perfect
entanglers constitute a polyhedron with seven faces, 
whose volume is exactly half that of the
tetrahedron. This implies that the among all the non-local two-qubit
operations, precisely half of them are capable of generating maximal
entanglement from some initially separable states.

For a two-qubit state $\psi$, define a quadratic function $\Ent
\psi=\psi^T P \psi$, where $P=-\frac12\sigma_y^1\sigma_y^2$ \cite{Makhlin:00}. 
It can be shown
that $\max_\psi |\Ent \psi|=\frac12$,  and
$\Ent \psi=0$ if and only if $\psi$ is an unentangled state.
The function $\Ent$ thus defines a measure of entanglement for a
pure state. If $|\Ent\psi|=\frac12$, we call $\psi$ a 
maximally entangled state.
It can be proved that the function $\Ent$ is invariant under
the local operations.

\begin{definition}[Perfect entangler]
  A two-qubit gate $U$ is called a perfect entangler if  it can
  produce a maximally entangled state from an unentangled one.
\end{definition}

\begin{definition}[Convex hull]
 The convex hull $C$ of $N$ points $p_1$, \dots, $p_N$ in $\R^n$
is given by
\begin{equation}
  \label{eq:65}
      C=\bigg\{\sum_{j=1}^N \theta_j p_j \ \big | \ \theta_j \ge 0 \text{ for all } j
    \text{ and } \sum_{j=1}^N \theta_j=1 \bigg\}.
  \end{equation}
\end{definition}

\begin{theorem} [Condition for perfect entangler]
\label{thm:1}
A two-qubit gate $U$ is a perfect entangler if and only if
 the convex hull of the
eigenvalues of $m(U)$ contains zero.
\end{theorem}

This result was first mentioned by Makhlin \cite{Makhlin:00} but no
proof was given.  We provide here a proof and then go on to develop a 
geometrical analysis that provides a quantification of the 
relative volume of perfect entanglers in $SU(4)$.

\begin{proof}
From the Cartan decomposition of $\su(4)$ in
Section~\ref{Sec:CartanSu4}, any two-qubit gate $U\in
U(4)$ can be written in the following form
\begin{equation}
  \label{eq:13}
U=e^{i\alpha}k_1Ak_2=e^{i\alpha} 
k_1\exp\{\frac{i}2(c_1\sigma_x^1\sigma_x^2 +c_2\sigma_y^1\sigma_y^2
+c_3\sigma_z^1\sigma_z^2)\} k_2,
\end{equation}
where $k_1$, $k_2\in {SU(2)\otimes SU(2)}$.
For any arbitrary unentangled state $\psi_0$, we have
\begin{equation}
  \label{eq:14}
    \Ent U\psi_0=\Ent e^{i\alpha}k_1Ak_2\psi_0=e^{i2\alpha}\Ent A\psi,
\end{equation}
where $\psi=k_2\psi_0$ is again an unentangled state. 
From Eq. (\ref{eq:14}), it is clear that
 $|\Ent U\psi_0|=|\Ent A\psi|$. Therefore, 
$U$ is a perfect entangler if and only if $A$ is a perfect entangler. 
Furthermore, we have
\begin{eqnarray}
  \label{eq:15}
  \aligned
\Ent A\psi&=\psi^TA^TPA\psi\\
&=(Q^\dag\psi)^T (Q^\dag
  AQ)^T(Q^TPQ)(Q^\dag AQ)(Q^\dag\psi)\\
&=\frac12 (Q^\dag\psi)^TF^2 (Q^\dag\psi),
\endaligned
\end{eqnarray}
where $Q$ and $F$ are defined as in Eq. (\ref{eq:4}) and (\ref{eq:5}),
respectively. The last equality in Eq. (\ref{eq:15}) holds since
$Q^TPQ=\frac12 I$. Let $\phi=Q^\dag\psi$.
  Since $\psi$ is an unentangled state, we get $\Ent \psi=0$. Hence,
  \begin{equation}
\label{eq:7}
\Ent\psi=\psi^TP\psi=\phi^TQ^TPQ\phi=\frac12\phi^T\phi=\frac12\big 
(\phi_1^2+\phi_2^2+\phi_3^2+\phi_4^2\big)=0.
  \end{equation}
Since $\psi^\dag\psi=1$, we have $\phi^\dag\phi=1$, that is,
\begin{equation}
\label{eq:8}
  |\phi_1|^2+|\phi_2|^2+|\phi_3|^2+|\phi_4|^2=1.
\end{equation}
Recall the definition of $F$ from Eq. (\ref{eq:5}):
\begin{equation}
  \label{eq:16}
F=\diag\{e^{i\frac{c_1-c_2+c_3}2}, e^{i\frac{c_1+c_2-c_3}2},
e^{-i\frac{c_1+c_2+c_3}2}, e^{i\frac{-c_1+c_2+c_3}2} \}.
\end{equation}
For simplicity, we denote the eigenvalues of $F$ as
$\{\lambda_k\}_{k=1}^4$. Then the eigenvalues of $m(U)$ are just 
$\{\lambda_k^2\}_{k=1}^4$. We have 
\begin{eqnarray}
  \label{eq:17}
    \Ent A\psi=\frac12 (Q^\dag\psi)^TF^2 (Q^\dag\psi)
=\frac12\phi^TF^2\phi
=\frac12\sum_{k=1}^4 \phi_k^2\lambda_k^2.
\end{eqnarray}
If $A$ is a perfect entangler, we have
\begin{eqnarray}
  \label{eq:10}
  \aligned
  \frac12&=|\Ent A\psi|=\frac12|\phi_1^2\lambda_1^2+\phi_2^2\lambda_2^2
+\phi_3^2\lambda_3^2+\phi_4^2\lambda_4^2|\\
&\le \frac12\big(|\phi_1^2\lambda_1^2|+|\phi_2^2\lambda_2^2|
+|\phi_3^2\lambda_3^2|+|\phi_4^2\lambda_4^2|\big)\\
&= \frac12\big(|\phi_1^2|+|\phi_2^2|+|\phi_3^2|+|\phi_4^2|\big)=\frac12.
\endaligned
\end{eqnarray}
The equality in Eq. (\ref{eq:10}) holds if and 
only if there exists a real number $\theta\in
 [0, 2\pi]$  such that
 \begin{equation}
   \label{eq:66}
     \phi_1^2\lambda_1^2=|\phi_1|^2e^{i2\theta}, \quad
  \phi_2^2\lambda_1^2=|\phi_2|^2e^{i2\theta}, \quad
 \phi_3^2\lambda_1^2=|\phi_3|^2e^{i2\theta},\quad
 \phi_4^2\lambda_1^2=|\phi_4|^2e^{i2\theta}.
\end{equation}
From Eq. (\ref{eq:7}), we obtain
\begin{equation}
  \label{eq:25}
  \phi_1^2+\phi_2^2+\phi_3^2+\phi_4^2=e^{i2\theta}\left(
\frac{|\phi_1|^2}{\lambda_1^2}+\frac{|\phi_2|^2}{\lambda_2^2}+
\frac{|\phi_3|^2}{\lambda_3^2}+\frac{|\phi_4|^2}{\lambda_4^2}\right)=0.
\end{equation}
Since $\frac{1}{\lambda_k}=\overline{\lambda}_k$, it follows that
\begin{equation}
  \label{eq:26}
  |\phi_1|^2\lambda_1^2+|\phi_2|^2\lambda_2^2
+|\phi_3|^2\lambda_3^2+|\phi_4|^2\lambda_4^2=0.
\end{equation}
From the relation in Eq. (\ref{eq:8}), we conclude that if $U$ is a perfect
entangler, 
the convex hull of the eigenvalues of $m(U)$ contains zero.

Conversely, suppose the convex hull of the
eigenvalues of $m(U)$ contains zero, that is, there exist
$\{\alpha_k\}_{k=1}^4 \subset [0, 1]$ such that
\begin{eqnarray}
  \label{eq:18}
\gathered
  \alpha_1^2\lambda_1^2+  \alpha_2^2\lambda_2^2 
+  \alpha_3^2\lambda_3^2+  \alpha_4^2\lambda_4^2=0,\\
  \alpha_1^2+\alpha_2^2+\alpha_3^2+  \alpha_4^2=1.
\endgathered
\end{eqnarray}
Let
\begin{equation}
  \label{eq:89}
\phi=\left(\frac{\alpha_1}{\lambda_1},
\frac{\alpha_2}{\lambda_2},\frac{\alpha_3}{\lambda_3},
\frac{\alpha_4}{\lambda_4}\right)^T,
\end{equation}
and $\psi=Q\phi$. From Eq. (\ref{eq:7}), we have
\begin{equation}
  \label{eq:90}
\Ent\psi=\frac12\phi^T\phi=
\frac12\left(\frac{\alpha_1^2}{\lambda_1^2}+
\frac{\alpha_2^2}{\lambda_2^2}+\frac{\alpha_3^2}{\lambda_3^2}+
\frac{\alpha_4^2}{\lambda_4^2}\right)=0.
\end{equation}
Hence $\psi$ is an unentangled state. From Eq. (\ref{eq:17}), we derive
\begin{equation}
  \label{eq:67}
    \Ent A\psi=\frac12 \phi^T F^2\phi=\frac12
\big(\alpha_1^2+\alpha_2^2+\alpha_3^2+  \alpha_4^2\big)=\frac12.
\end{equation}
Therefore, $U$ is a perfect entangler.
\end{proof}

We now derive the conditions under which points $[c_1, c_2, c_3]$ in the Weyl chamber
$\a^+$ are perfect entanglers.  We begin with two corollaries to
Theorem~\ref{thm:1}.
\begin{corollary}
  \label{coro:1}
If $[c_1, c_2, c_3]$ is a perfect entangler,
then $[\pi-c_1, c_2, c_3]$ and
$[\frac{\pi}2-c_1,\frac{\pi}2-c_2,\frac{\pi}2-c_3]$
 are both perfect entanglers.
\end{corollary}
\begin{proof}
We know that $[c_1, c_2, c_3]$ and $[c_1, -c_2, -c_3]$
correspond to the same two-qubit gate. Since the 3-Torus has the
minimum positive 
period $\pi$, $[-\pi+c_1, -c_2, -c_3]$ also belongs to the same local
equivalence class.
 From Eq. (\ref{eq:25}) and (\ref{eq:26}), if $[c_1, c_2, c_3]$ is a
  perfect entangler, so is $[-c_1, -c_2, -c_3]$. Therefore,
  $[\pi-c_1, c_2, c_3]$ is a perfect entangler.

From Theorem \ref{thm:1}, $U$ is a perfect entangler if and only
if the convex hull of the eigenvalues of $m(U)$ contains zero,
 that is, there exist
$\{\alpha_k\}_{k=1}^4 \subset [0, 1]$ such that
\begin{align}
  \label{eq:23}
  \alpha_1^2e^{i(c_1-c_2+c_3)}+\alpha_2^2 e^{i(c_1+c_2-c_3)}
+  \alpha_3^2e^{-i(c_1+c_2+c_3)}+  \alpha_4^2 e^{i(-c_1+c_2+c_3)}=0 ,\\
\label{eq:24}
  \alpha_1^2+\alpha_2^2+\alpha_3^2+  \alpha_4^2=1.
\end{align}
Substitute the coordinates of the 
point $[\frac{\pi}2-c_1,\frac{\pi}2-c_2,\frac{\pi}2-c_3]$ 
into Eq. (\ref{eq:23}):
\begin{eqnarray}
  \label{eq:22}
i\big\{\alpha_1^2e^{-i(c_1-c_2+c_3)}+\alpha_2^2 e^{-i(c_1+c_2-c_3)}
+\alpha_3^2e^{i(c_1+c_2+c_3)}+  \alpha_4^2 e^{-i(-c_1+c_2+c_3)}\big\}=0.
\end{eqnarray}
Together with Eq. (\ref{eq:24}), it is clear that  
$[\frac{\pi}2-c_1,\frac{\pi}2-c_2,\frac{\pi}2-c_3]$ is a perfect entangler.
\end{proof}

\begin{corollary}
\label{coro:2}
For a two-qubit gate $U$, if its corresponding point in the Weyl
chamber $\a^+$ is $[c_1, \frac{\pi}2-c_1, c_3]$, 
$[c_1, c_1-\frac{\pi}2, c_3]$, or $[c_1, c_2,  \frac{\pi}2-c_2]$, 
$U$ is a perfect entangler.
\end{corollary}
\begin{proof}
For the gate $[c_1, \frac{\pi}2-c_1, c_3]$, the eigenvalues of $m(U)$ are
\begin{eqnarray}
\label{eq:coro2}
\aligned
 \, &\{e^{i(c_1-c_2+c_3)}, e^{i(c_1+c_2-c_3)},
e^{-i(c_1+c_2+c_3)}, e^{i(-c_1+c_2+c_3)} \}\\
&=e^{-i(c_1+c_2+c_3)}\, \{e^{i2(c_1+c_3)}, e^{i2(c_1+c_2)},
1, e^{i2(c_2+c_3)} \}.
\endaligned
\end{eqnarray}
The convex hull of the eigenvalues of $m(U)$ always contains the
origin, and thus  $[c_1, \frac{\pi}2-c_1, c_3]$ is a perfect
entangler. The other cases can be proved similarly.
\end{proof}

Note that for $[c_1, \frac{\pi}2-c_1, c_3]$, picking
 $c_1=\frac{\pi}2$ and $c_3=0$, we obtain
the perfect entangler [CNOT]; picking $c_1=\frac{\pi}4$ and
$c_3=\frac{\pi}4$, we get the perfect entangler $[\sqrt{\text{SWAP}}]$.

With these corollaries in hand, we can proceed to 
derive the conditions under which a 
general point $[c_1, c_2, c_3]$ on the
3-Torus is a perfect entangler.

\begin{theorem}[Perfect entangler on 3-Torus]
  \label{thm:2}
Consider a two-qubit gate $U$ and its corresponding representation
$[c_1, c_2, c_3]$ on the 3-Torus. $U$ is a perfect entangler if and
only if one of the following two conditions is satisfied:
\begin{eqnarray}
\aligned  
  \label{eq:27}
\frac\pi{2}\le c_i+c_k\le c_i+c_j+\frac\pi{2} \le \pi,\\
\frac{3\pi}2 \le c_i+c_k\le c_i+c_j+\frac\pi{2} \le 2\pi,
\endaligned
\end{eqnarray}
where $(i, j, k)$ is a permutation of $(1, 2, 3)$.
\end{theorem}

 \begin{figure}[tb]
 \begin{center}
 \begin{tabular}{cc}
 \footnotesize
  \psfrag{x}[][]{$e^{i2(c_j+c_k)}$}
  \psfrag{y}[][]{$e^{i2(c_i+c_j)}$}
  \psfrag{z}[][]{$e^{i2(c_i+c_k)}$}
\includegraphics[width=0.3\hsize]{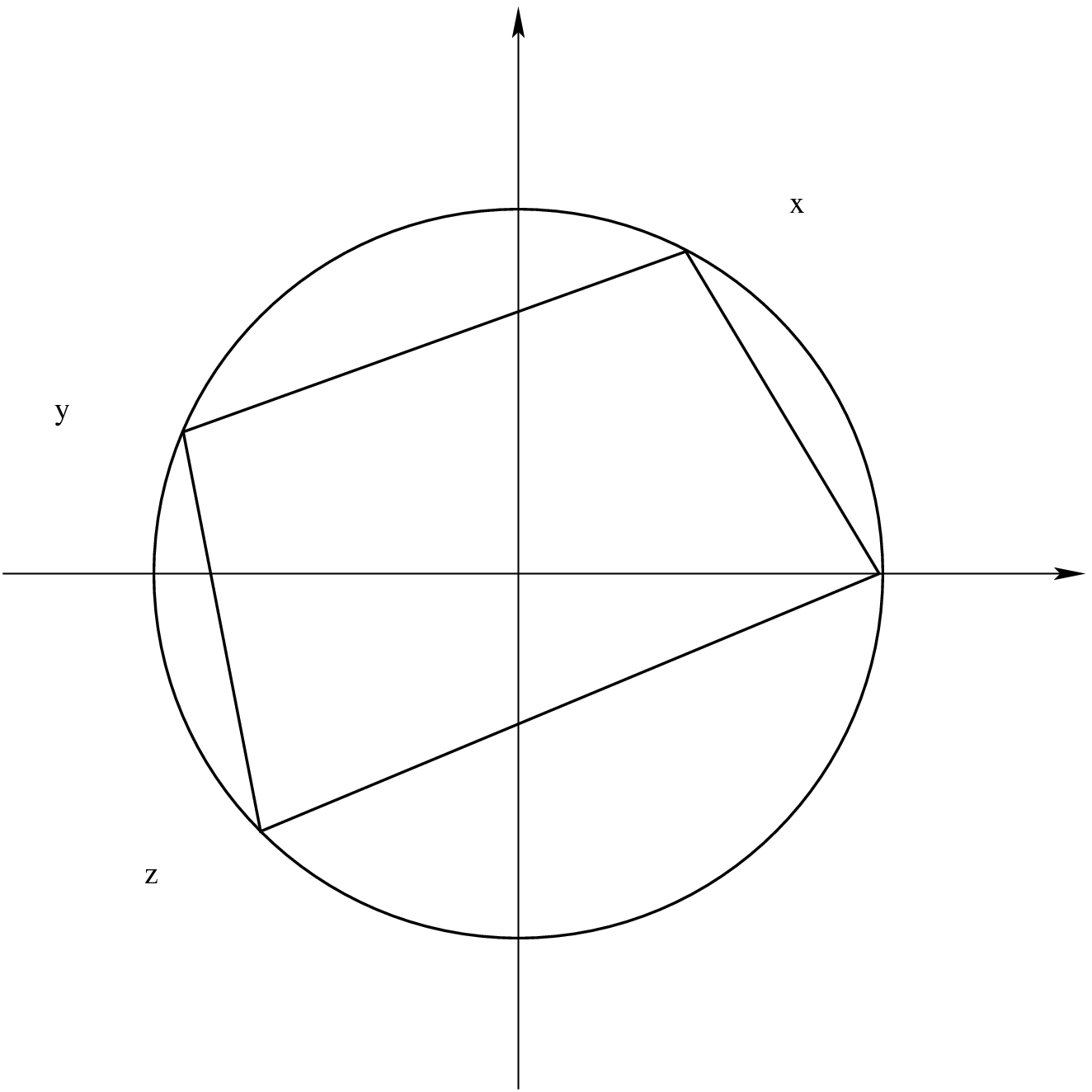} 
&
 \footnotesize
 \psfrag{x}[][]{$e^{i2(c_j+c_k)}$}
  \psfrag{y}[][]{$e^{i2(c_i+c_j)}$}
  \psfrag{z}[][]{$e^{i2(c_i+c_k)}$}
 \includegraphics[width=0.3\hsize]{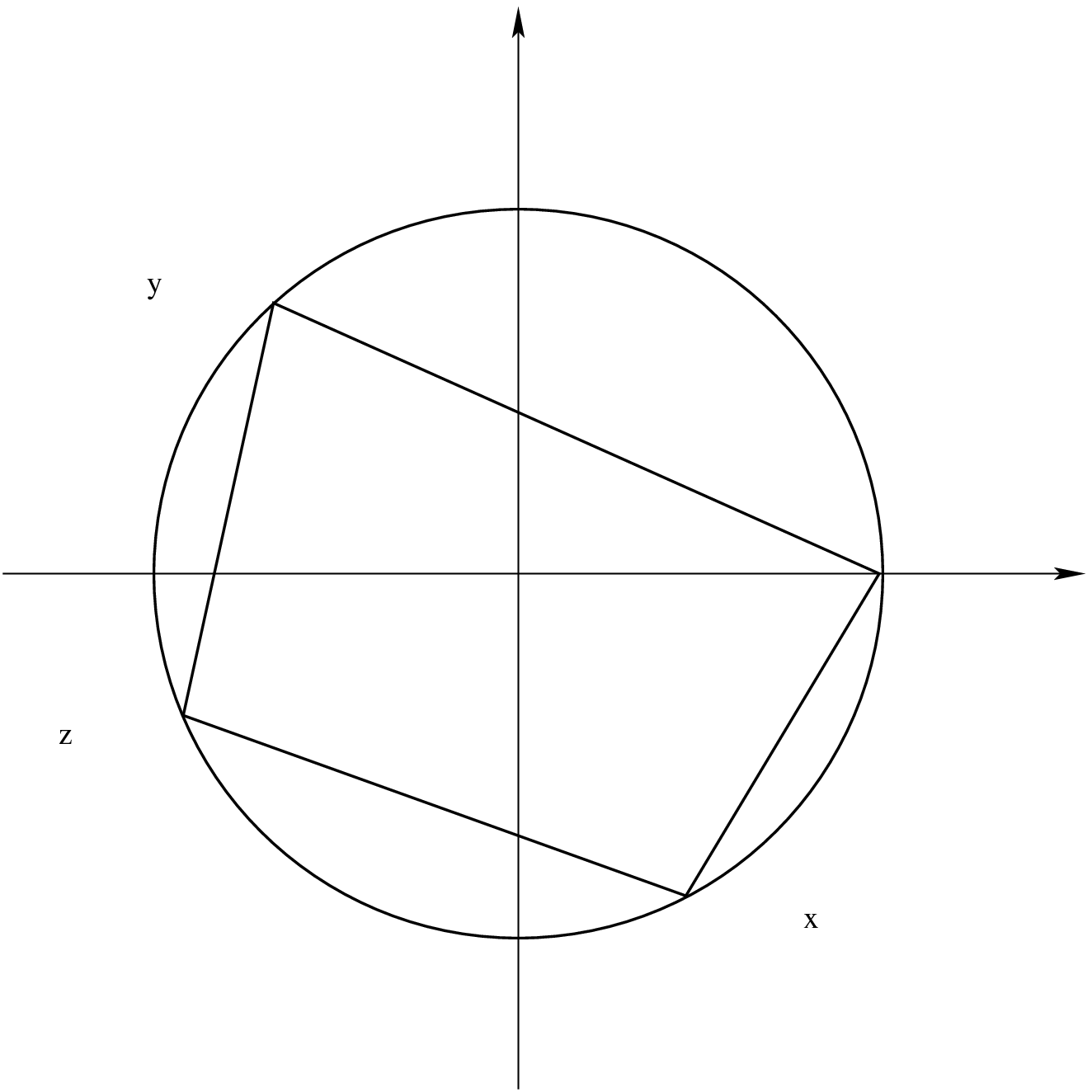} \\
 (A) & (B)
 \end{tabular} 
 \end{center}
 \caption{Illustration of the proof of Theorem~\ref{thm:2}.}
 \label{fig:1}
\end{figure}

\begin{proof}
Given the eigenvalues of $m(U)$ in Eq. (\ref{eq:coro2}),
it suffices to study whether the convex hull of
$\{1, e^{i2(c_1+c_2)}, e^{i2(c_1+c_3)} e^{i2(c_2+c_3)} \}$ 
 contains the origin or not. 
Suppose that one of the conditions in Eq. (\ref{eq:27}) is
 satisfied. In this case the points  $\{e^{i2(c_1+c_2)}, e^{i2(c_1+c_3)},
e^{i2(c_2+c_3)} \}$ have to be on the unit
 circle as shown in Figure~\ref{fig:1} (A) or (B). It is clear that
the convex hull of these three points contains the origin.
From Theorem~\ref{thm:1}, $U$ is therefore a perfect entangler.

Conversely, suppose that $U$ is a perfect entangler. Then the convex hull of
 the eigenvalues of $m(U)$ contains the origin. If all the three points
 $\{e^{i2(c_1+c_2)}, e^{i2(c_1+c_3)}, e^{i2(c_2+c_3)} \}$ are on
 the upper or lower semi-circle, the convex hull of  $\{1,
 e^{i2(c_1+c_2)}, e^{i2(c_1+c_3)}, e^{i2(c_2+c_3)} \}$ does not
 contain the origin. Therefore, we can always pick one point
 on the upper semi-circle and one point on the lower semi-circle such
 that one of the two conditions in Eq. (\ref{eq:27}) is
 satisfied.
\end{proof}

The above analysis shows that whether a two-qubit gate is a
perfect entangler or not is only determined by its geometric representation 
$[c_1, c_2, c_3]$ on the 3-Torus. Recall that in
Section~\ref{sec:geo}, we show that the
local equivalence classes of two-qubit gates are in one-to-one
correspondence with the points of the Weyl chamber $\a^+$, which
can be represented by a tetrahedron as shown in Figure~\ref{fig:2}(C).
We are now ready for the final stage of the procedure, namely to identify those points in 
the {\em tetrahedron}
that correspond to perfect entanglers.

\begin{figure}[t]
\begin{center}
 \footnotesize
 \psfrag{N}[][]{$N$}
 \psfrag{M}[][]{$M$}
 \psfrag{L}[][]{$L$}
 \psfrag{P}[][]{$P$}
 \psfrag{Q}[][]{$Q$}
 \psfrag{A1}[][]{$A_1$}
 \psfrag{A2}[][]{$A_2$}
 \psfrag{A3}[][]{$A_3$}
 \psfrag{O}[][]{$O$}
\includegraphics[width=0.4\hsize]{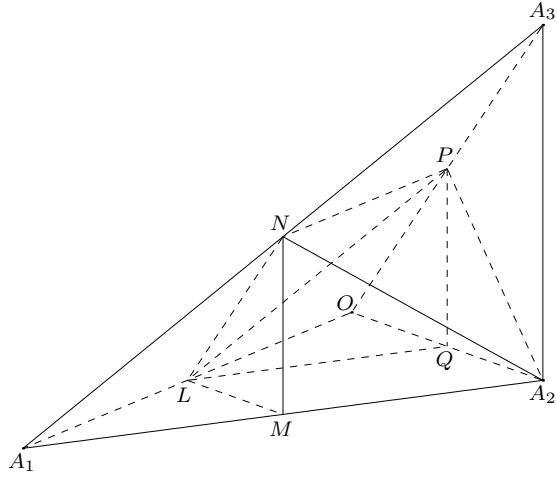} 
\end{center}
 \caption{Polyhedron $\text{LMNPQA}_2$ corresponds to 
perfect entanglers in the Weyl chamber $\a^+$ (see Figure~\ref{fig:2}
(C)), where L, M, N, P, and Q are the
midpoints of the line segments $\text{A}_1\text{Q}$,
$\text{A}_1\text{A}_2$, $\text{A}_1\text{A}_3$, 
$\text{A}_3\text{Q}$, and $\text{A}_2\text{Q}$, respectively.
$\text{P}$ corresponds to the gate $\sqrt{\text{SWAP}}$, N to
its inverse, and $L$ to the CNOT gate.}
 \label{fig:pe}
\end{figure}

Consider a two-qubit gate $[c_1, c_2, c_3]$.
As shown in Figure~\ref{fig:pe}, in the tetrahedron
$\text{OA}_1\text{A}_2\text{A}_3$, we have $c_1\ge c_2 \ge
c_3\ge 0$. Hence  $2(c_1+c_2)\ge 2(c_1+c_3) \ge 2(c_2+c_3)\ge 0$.
As in the proof of Theorem~\ref{thm:2}, 
consider the convex hull of $\{1, e^{i2(c_1+c_2)}, e^{i2(c_1+c_3)}
e^{i2(c_2+c_3)} \}$. We can identify the following 
three cases of the gates that do not 
provide maximal entanglement, and are thus not perfect entanglers:
\begin{itemize}
\item If $c_1+c_2\le \frac\pi{2}$, that is, 
all the $ e^{i2(c_j+c_k)}$ are on the upper
semi-circle, the gate is not a perfect
entangler. In the tetrahedron $\text{OA}_1\text{A}_2\text{A}_3$,
$c_1+c_2\le \frac\pi{2}$ corresponds to the tetrahedron $\text{LQPO}$.
\item If $c_2+c_3\ge \frac\pi{2}$, that is, 
all the $ e^{i2(c_j+c_k)}$ are on the lower
semi-circle, the gate is not a perfect
entangler either. This case corresponds to the tetrahedron
$\text{NPA}_2\text{A}_3$. 
\item From Theorem~\ref{thm:2}, we obtain that the gates represented by points in the
set $\{X\in \a^+ |\,2(c_1+c_3)\ge 2(c_2+c_3)+\pi \}$
 are not perfect entanglers. This set is the tetrahedron
$\text{LMNA}_1$.
\end{itemize}
The set of perfect entanglers can thus be obtained by removing these three
tetrahedra from $\text{OA}_1\text{A}_2\text{A}_3$. 
This is done in Figure~\ref{fig:pe} where it is thereby evident that the polyhedron 
 $\text{LMNPQA}_2$ is the residual set of perfect entanglers.  Here
the point $\text{P}$ corresponds to the gate $\sqrt{\text{SWAP}}$, N
to its inverse, and $L$ to the CNOT gate.
Computing the volume of the Weyl chamber $\text{O}\text{A}_1\text{A}_2\text{A}_3$
and of these three polyhedra, we have
\begin{eqnarray}
  \label{eq:76}
  \gathered
  V(\text{OA}_1\text{A}_2\text{A}_3)=\frac{\pi^3}{24},\\
V(\text{LQPO})=\frac{\pi^3}{192},\quad
V(\text{NPA}_2\text{A}_3)=\frac{\pi^3}{96},\quad
V(\text{LMNA}_1)=\frac{\pi^3}{192}.
\endgathered
\end{eqnarray}
Therefore, the volume of the polyhedron $\text{LMNPQA}_2$ is
$\pi^3/48$, which
is half of the volume of $\text{OA}_1\text{A}_2\text{A}_3$. 
This implies that
among all the non-local two-qubit gates, half of them are perfect entanglers.
Note that the  polyhedron $\text{LMNPQA}_2$ is symmetric with respect to
the plane $c_1=\pi/2$, which provides a geometric explanation of
Corollary~\ref{coro:1}. The points in Corollary~\ref{coro:2}
correspond to the triangles LMN, LPQ, 
and $\text{NPA}_2$, which are three faces of
the set of perfect entanglers. Also recall that the line $\text{OL}$
represents all the Controlled-$U$ gates. Hence
 CNOT, located at $\text{L}$, is the only Controlled-$U$ gate that is a perfect entangler.
Thus we see that the geometric
representation provides an intuitive visual picture to understand the
non-local properties of two-qubit gates,
as well as allowing quantification of the weight of perfect entanglers.

\section{Physical generation of non-local gates}
We now investigate the universal quantum computation and
simulation potential of a given physical Hamiltonian. 
We first study the gates that can
be generated by a Hamiltonian directly. Generally speaking, these
gates form a one dimensional subset on the 3-Torus geometric
representation of non-local gates. For any arbitrary two-qubit
gate, we will explicitly construct a quantum circuit that can simulate it
exactly with a guaranteed small number of operations. Construction of efficient circuits 
is especially important in the theoretical 
design and experimental implementations 
of quantum information processing. 
We assume only that we can turn on local operations individually.
Our starting point is thus any arbitrary single qubit 
operation and a two-body interaction Hamiltonian.
The single and two-qubit operations may be, for example, a 
sequence of pulses of an optical field that are suitably tuned 
and focused on each individual qubit.  The qubits may be represented by either a solid state system such as a quantum dot 
in a cavity \cite{Imamoglu:99}, or by a gas phase system such as an optical lattice \cite{Deutsch:98}.

\subsection{Non-local operations generated by a given Hamiltonian}
\label{sec:nonlocal}
In this subsection we investigate the non-local gates 
that can be generated by a given
Hamiltonian $H$ for a time duration $t$, 
that is, $U(t)=\exp iHt$. Recall that $\k$ in Eq. (\ref{eq:29}) 
is the Lie subalgebra corresponding to $K$, the Lie subgroup of all the
local gates.
Therefore, $\k$ can be viewed as the local part in $\su(4)$, 
and $\p$ as the
non-local part. If the Hamiltonian $H$ contains the non-local part, that
is, $iH\cap \p\neq \emptyset$, then $H$ can generate 
non-local gates.

We first consider a Hamiltonian $H$ for which $iH$ is in the Cartan
subalgebra $\a$, and then extend to the general case.
Assume $H=\frac12(c_1\sigma_x^1\sigma_x^2+c_2\sigma_y^1\sigma_y^2
+c_3\sigma_z^1\sigma_z^2)$.
The  local equivalence classes of
$U(t)$ form a continuous flow on the 3-Torus
as time evolves. This provides us a geometric 
picture to study the
properties of the gates generated by a given Hamiltonian. To
illustrate the ideas, we consider the following examples.

\begin{example}[Exchange Hamiltonians]
\noindent{\bf{(1) Isotropic (Heisenberg) exchange:}
 $H_1=\frac14(
\sigma_x^1\sigma_x^2+\sigma_y^1\sigma_y^2+\sigma_z^1\sigma_z^2) $}

In this case, the two-qubit gate $U(t)$ generated by the Hamiltonian
$H_1$ is 
\begin{eqnarray}
  \label{eq:77}
   U(t)=\exp{iHt}= \exp i\frac{t}4 \big(\sigma_x^1\sigma_x^2+
\sigma_y^1\sigma_y^2+\sigma_z^1\sigma_z^2\big).
\end{eqnarray}
Hence the Hamiltonian $H_1$ generates the flow $[\frac{t}2,
\frac{t}2, \frac{t}2]$ on the 3-Torus. 
The local invariants can thus be computed from Eq. (\ref{eq:11}): 
\begin{eqnarray}
  \label{eq:78}
  \aligned
  G_1(t)&=\frac{\tr^2(m)}{16\det U}=(\cos^3\frac{t}2-i\sin^3\frac{t}2)^2
=\frac{e^{it}}{16}(3+e^{-2it})^2,\\
G_2(t)&=\frac{\tr^2(m)-\tr(m^2)}{4\det U}
=4(\cos^6 \frac{t}2-\sin^6\frac{t}2)-\cos^3t=3\cos t.
\endaligned
\end{eqnarray}
We reduce the symmetry of the 
flow to the Weyl chamber $\a^+$, as shown in Figure~\ref{fig:pe}.
We obtain that for $t\in [2k\pi, 2k\pi+\pi]$, the trajectory is  $[\frac{t}2,
\frac{t}2, \frac{t}2]$;
and for  $t\in [2k\pi+\pi, 2(k+1)\pi]$, the trajectory is $[\frac{t}2,
\pi-\frac{t}2, \pi-\frac{t}2]$.
Therefore, the flow generated by the isotropic
Hamiltonian $H_1$ evolves along $\text{OA}_3\text{A}_1$, which
corresponds to all the local equivalence classes that can be generated by
$H_1$. Moreover, it can easily be seen that $\sqrt{\text{SWAP}}$ and its inverse are
 the only two perfect entanglers that can
be achieved by this Hamiltonian.

\noindent{\bf{(2) Two-dimensional exchange, {\it i.e.}, XY Hamiltonian:}
    $H_2=\frac14(\sigma_x^1\sigma_x^2+\sigma_y^1\sigma_y^2)$}

The Hamiltonian  $H_2$ generates the flow $[\frac{t}2,
\frac{t}2, 0]$ for $t\in [2k\pi, 2k\pi+\pi]$,
 and $[\frac{t}2, \pi-\frac{t}2, 0]$ for  $t\in [2k\pi+\pi, 2(k+1)\pi]$.
Hence the trajectory evolves along $\text{OA}_2\text{A}_1$. It
is evident that $H_2$ can generate a set of perfect entanglers that
corresponds to the line segments $\text{QA}_2$ and $\text{A}_2\text{M}$ 
in $\a^+$. Note that $\text{A}_2\text{M}$ represents exactly the same
local equivalence classes as $\text{QA}_2$.
The local invariants of $U(t)$ are
$G_1(t)=\cos^4\frac{t}2$ and $G_2(t)=1+2\cos t$.

\noindent{\bf{(3) One-dimensional exchange, {\it i.e.}, Ising Hamiltonian:}
 $H_3=\frac14\sigma_y^1\sigma_y^2$}

 The trajectory  generated by the 
Hamiltonian  $H_3$ in $\a^+$ is  $[\frac{t}2,
0, 0]$, which evolves along the line  $\text{OA}_1$. Hence the gates
generated by the Hamiltonian $H_3$ are all the
Controlled-$U$ gates. As noted above, CNOT, located at $\text{L}$, is the
only perfect entangler that can be generated by this Hamiltonian.
The local invariants of $U(t)$ are $G_1(t)=\cos^2\frac{t}2$ and
$G_2(t)=2+\cos t$.
\end{example}

For any arbitrary $H=\frac12(c_1\sigma_x^1\sigma_x^2
+c_2\sigma_y^1\sigma_y^2+c_3\sigma_z^1\sigma_z^2)$,  the trajectory on
the 3-Torus is $[c_1t, c_2t, c_3t]$. If both $c_1/c_2$
and $c_1/c_3$ are rational, the trajectory generated by the
Hamiltonian $H$ forms a loop on the 3-Torus. If either $c_1/c_2$ or
$c_1/c_3$ is irrational, the trajectory forms 
a proper dense subset of 3-Torus.

Next let us consider the case when $iH\in \p$. Recall that 
we have $\p=\cup_{k\in K}\Ad_k(\a)$. Hence for any arbitrary $iH\in
\p$, there exists a local gate $k\in SU(2)\otimes SU(2)$ such that 
\begin{equation}
  \label{eq:2}
  \Ad_{k}(iH)=iH_a,
\end{equation}
where $H_a=\frac{1}2(c_1\sigma_x^1\sigma_x^2
+c_2\sigma_y^1\sigma_y^2+c_3\sigma_z^1\sigma_z^2)$.  It follows that 
\begin{equation}
  \label{eq:68}
  U(t)=\exp(iHt)=\exp\big(k^\dag iH_a k t\big)=
k^\dag\exp\big( \frac{i}2(c_1\sigma_x^1\sigma_x^2
+c_2\sigma_y^1\sigma_y^2+c_3\sigma_z^1\sigma_z^2)t \big)k.
\end{equation}
Therefore, the trajectory of $U(t)$ in the Weyl chamber $\a^+$ is
 $[c_1t, c_2t, c_3t]$. Eq. (\ref{eq:2}) also implies that $H$ and $H_a$
 have the same set of eigenvalues. We can thus use this property to derive
 the triplet  $[c_1, c_2, c_3]$ explicitly. The following
 example shows how to find the flow in the Weyl chamber $\a^+$ for a
 given Hamiltonian $H$ with $iH\in \p$.
 \begin{example}[Generalized exchange with cross-terms]
   Consider the generalized anisotropic exchange Hamiltonian
   $H=\frac12(J_{xx}\sigma_x^1\sigma_x^2+J_{yy}\sigma_y^1\sigma_y^2
+J_{xy}\sigma_x^1\sigma_y^2+J_{yx}\sigma_y^1\sigma_x^2)$
discussed   in~\cite{Vala:02b}. The eigenvalues of $H$ are
   \begin{eqnarray}
     \label{eq:39}
\gathered
 \frac12\bigg\{\sqrt{(J_{xx}+J_{yy})^2+(J_{xy}-J_{yx})^2}, \
-\sqrt{(J_{xx}+J_{yy})^2+(J_{xy}-J_{yx})^2},\\
\qquad \sqrt{(J_{xx}-J_{yy})^2+(J_{xy}+J_{yx})^2},\
-\sqrt{(J_{xx}-J_{yy})^2+(J_{xy}+J_{yx})^2} \ \bigg\},
\endgathered
   \end{eqnarray}
whereas the eigenvalues of $H_a$ are
\begin{equation}
  \label{eq:75}
  \frac12 \{ -c_1+c_3+c_2, -c_1-c_3-c_2, c_1+c_3-c_2, c_1-c_3+c_2\}.
\end{equation}
Since $H$ and $H_a$ have the same set of eigenvalues, by comparing
Eqs. (\ref{eq:39}) and (\ref{eq:75}) and recalling that $c_1\ge c_2
\ge c_3\ge 0$, we find
\begin{eqnarray}
  \label{eq:84}
  \aligned
c_1&=\frac12\bigg(\sqrt{(J_{xx}+J_{yy})^2+(J_{xy}-J_{yx})^2}
+\sqrt{(J_{xx}-J_{yy})^2+(J_{xy}+J_{yx})^2}\bigg)\\
c_2&=\frac12\bigg|\sqrt{(J_{xx}+J_{yy})^2+(J_{xy}-J_{yx})^2}
-\sqrt{(J_{xx}-J_{yy})^2+(J_{xy}+J_{yx})^2}\bigg|\\
c_3&=0.
\endaligned
\end{eqnarray}
Therefore, the flow generated by this Hamiltonian in the Weyl chamber
$\a^+$ is $[c_1t, c_2t, 0]$, which evolves in the plane
$\text{OA}_1\text{A}_2$. 
 \end{example}

Now we consider the general case when $iH\in \su(4)$ and $H$ contains
 both the local and non-local part.
 To derive the trajectory of $U(t)=\exp iHt$ on the 3-Torus, we
 first compute the local invariants of $U(t)$ as in Eqs. (\ref{eq:88})
 and (\ref{eq:32}):
 \begin{eqnarray}
   \label{eq:79}
   \aligned
  G_1(t)&=\frac{\tr^2\big(m(U(t))\big)}{16},\\
G_2(t)&=\frac{\tr^2\big(m(U(t))\big)-\tr\big(m^2(U(t))\big)}{4}.
\endaligned
\end{eqnarray}
Then from the relation of the local invariants and $c_i$, 
we can obtain the flow $[c_1(t), c_2(t), c_3(t)]$ on
the 3-Torus by solving Eq. (\ref{eq:11}):
\begin{eqnarray}
  \label{eq:43}
    \aligned
G_1&=\cos^2 c_1\cos^2 c_2\cos^2 c_3-\sin^2 c_1\sin^2 c_2\sin^2 c_3
+\frac{i}4 \sin 2c_1\sin 2c_2\sin 2c_3\\
G_2&=4\cos^2 c_1\cos^2 c_2\cos^2 c_3-4\sin^2 c_1\sin^2
c_2\sin^2 c_3-\cos 2c_1\cos 2c_2\cos 2c_3.
\endaligned
\end{eqnarray}

\begin{example} [Josephson junction charge-coupled qubits]
For Josephson (charged-coupled) qubits~\cite{Makhlin:99},  elementary two-qubit gates are
  generated by the Hamiltonian $H_J=-\frac12
  E_J(\sigma_x^1+\sigma_x^2)+(E_J^2/E_L)\sigma_y^1\sigma_y^2 $. If
  $E_J$ is tuned to $\alpha E_L$, $\alpha\in \R$,
the local invariants can be obtained
  \begin{eqnarray}
    \label{eq:6}
\aligned
G_1&=\frac1{(1+\alpha^2)^2}\big(\alpha^2(x^2+y^2-1)+x^2 \big)^2,\\
G_2&=\frac1{1+\alpha^2}\big(3\alpha^2-1-4y^2\alpha^2+
8\alpha^2x^2y^2+4x^2-4x^2\alpha^2\big),
\endaligned
  \end{eqnarray}
where
\begin{equation}
  \label{eq:19}
  x=\cos{\alpha^2E_L} t, \quad y=\cos \sqrt{(\alpha^2+1)}\alpha E_Lt.
\end{equation}
By solving Eq. (\ref{eq:43}), we find that the flow 
generated by this Hamiltonian on the 3-Torus is:
\begin{eqnarray}
  \label{eq:41}
  \aligned
c_1(t)&=\alpha^2E_L t-\omega(t),\\
c_2(t)&=\alpha^2E_L t+\omega(t),\\
c_3(t)&=0,
\endaligned
\end{eqnarray}
where $\omega(t)=\tan^{-1}\sqrt{(1+\alpha^2y^2)/(\alpha^2-\alpha^2y^2)}$.
Since $c_3=0$, the Hamiltonian $H_J$ can reach only those local
equivalence classes on the base $\text{OA}_1\text{A}_2$, as shown in
Figure~\ref{fig:pe}. Therefore, the Hamiltonian $H_J$ is not able to
generate the perfect entangler $[\sqrt{\text{SWAP}}]$.
The trajectory generated by $H_J$ in the Weyl chamber $\a^+$ is shown
in Figure~\ref{fig:4}.

For the Hamiltonian $H_J$ to achieve [CNOT], we need to
solve Eq. (\ref{eq:6}) for $G_1=0$ and $G_2=1$. After some
algebraic derivations, we find
\begin{eqnarray}
  \label{eq:21}
  x^2&=\frac12, \quad
y^2&=\frac{\alpha^2-1}{2\alpha^2}.
\end{eqnarray}
It follows that
\begin{eqnarray}
  \label{eq:40}
  \gathered
t=\frac{(2k+1)\pi}{4\alpha^2E_L},\\
2\alpha^2\cos{\sqrt{1+\alpha^{-2}}\frac{2k+1}4 \pi}=\alpha^2-1,
\endgathered
\end{eqnarray}
where $k\in \Z$. When $E_L=1$, numerical solution of these expressions
shows that the minimum time solution  for the Hamiltonian $H_J$
to achieve [CNOT] is obtained for $\alpha=1.1991$, and 
the minimum time is $2.7309$.

\begin{figure}[tb]
\begin{center}
\begin{tabular}{cc}
 \footnotesize
 \psfrag{A}[][]{$O$}
 \psfrag{B}[][]{$\pi/4$}
 \psfrag{C}[][]{$\pi/2$}
 \psfrag{D}[][]{$3\pi/4$}
 \psfrag{E}[][]{$\pi$}
 \psfrag{GG}[][]{$\frac{\pi}4$}
 \psfrag{GH}[][]{$\frac{\pi}2$}
 \psfrag{HG}[][]{$\frac{3\pi}4$}
 \psfrag{HH}[][]{$\pi$}
 \psfrag{C1}[][]{$c_1$}
\psfrag{C2}[][]{$c_2$}
\includegraphics[width=0.3\hsize]{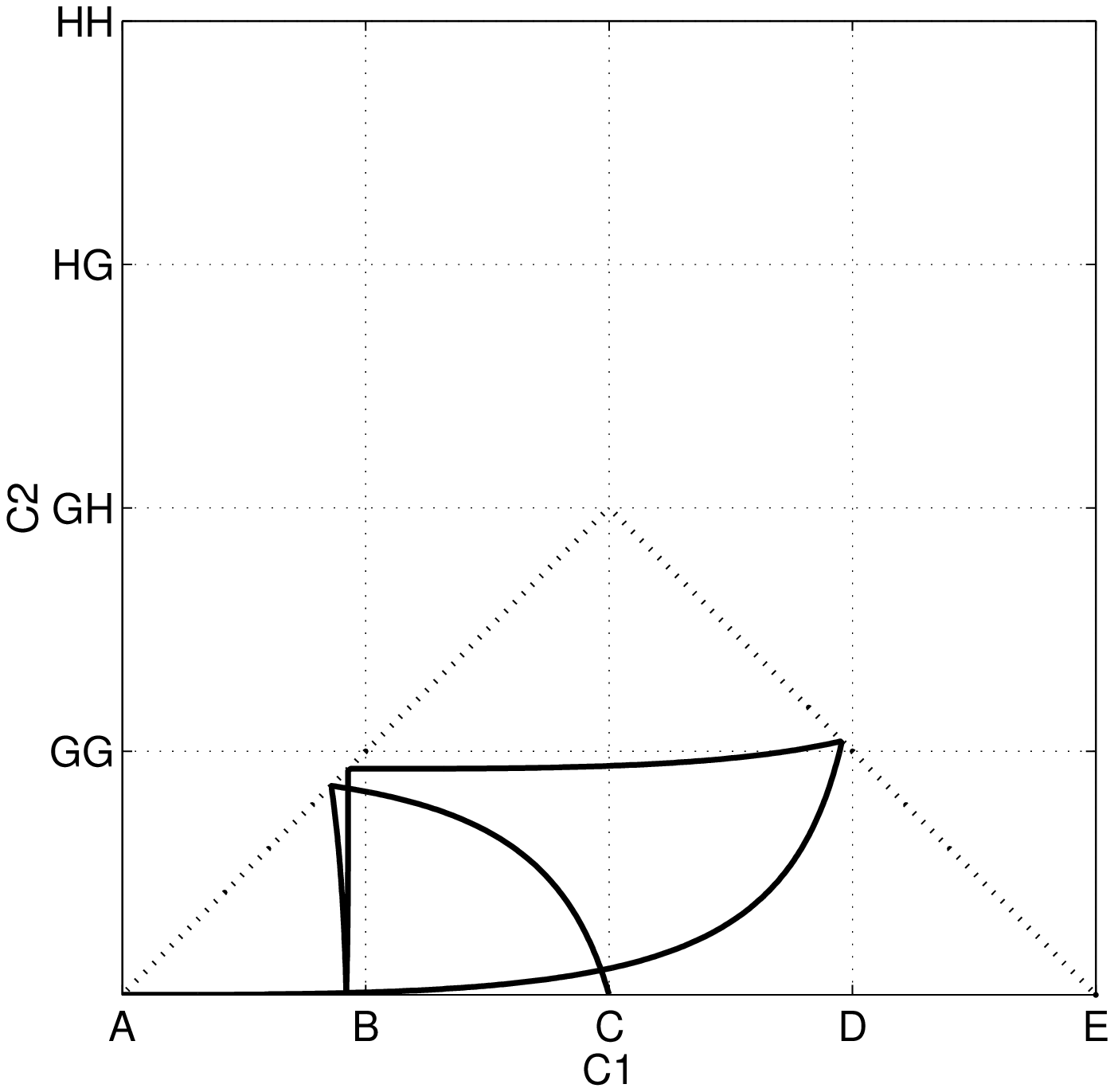}
&
 \footnotesize
 \psfrag{A}[][]{$O$}
 \psfrag{B}[][]{$\pi/4$}
 \psfrag{C}[][]{$\pi/2$}
 \psfrag{D}[][]{$3\pi/4$}
 \psfrag{E}[][]{$\pi$}
 \psfrag{GG}[][]{$\frac{\pi}4$}
 \psfrag{GH}[][]{$\frac{\pi}2$}
 \psfrag{HG}[][]{$\frac{3\pi}4$}
 \psfrag{HH}[][]{$\pi$}
 \psfrag{C1}[][]{$c_1$}
\psfrag{C2}[][]{$c_2$}
\includegraphics[width=0.3\hsize]{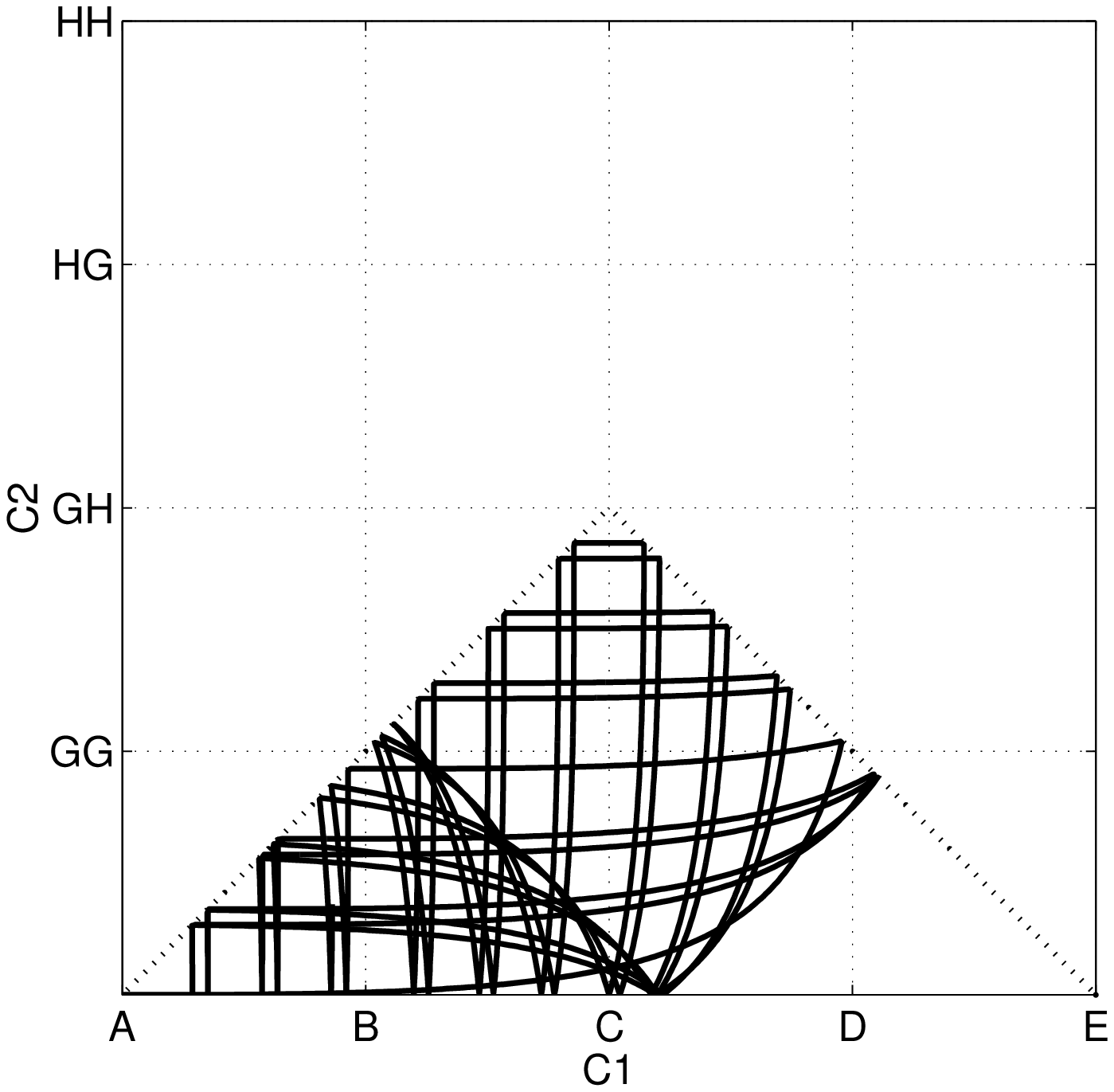}\\
$\alpha=1.1991$, $t=2.7309$ & $\alpha=1.1991$, $t=20$\\
 \footnotesize
 \psfrag{A}[][]{$O$}
 \psfrag{B}[][]{$\pi/4$}
 \psfrag{C}[][]{$\pi/2$}
 \psfrag{D}[][]{$3\pi/4$}
 \psfrag{E}[][]{$\pi$}
 \psfrag{GG}[][]{$\frac{\pi}4$}
 \psfrag{GH}[][]{$\frac{\pi}2$}
 \psfrag{HG}[][]{$\frac{3\pi}4$}
 \psfrag{HH}[][]{$\pi$}
 \psfrag{C1}[][]{$c_1$}
\psfrag{C2}[][]{$c_2$}
\includegraphics[width=0.3\hsize]{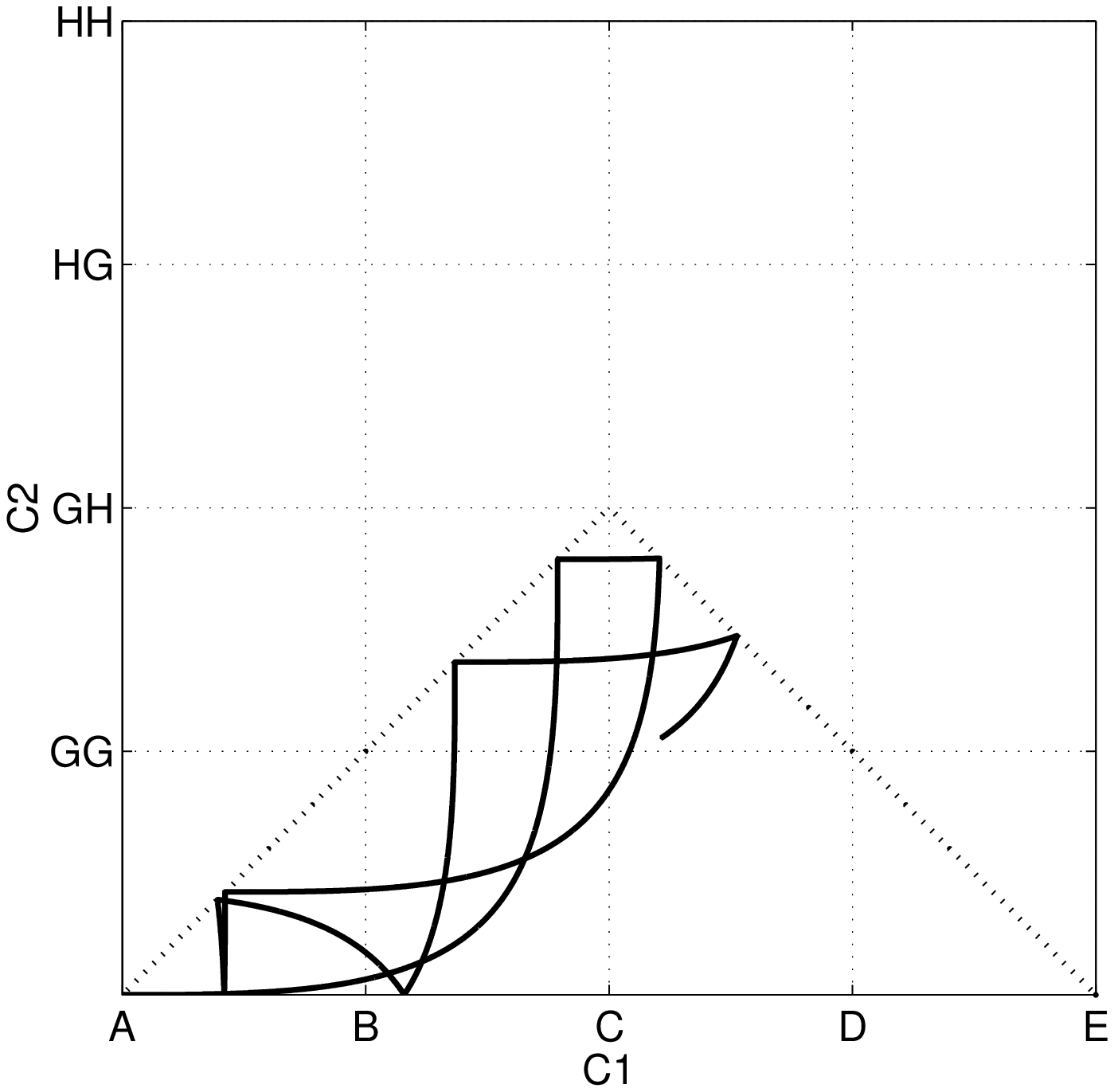}
&
 \footnotesize
 \psfrag{A}[][]{$O$}
 \psfrag{B}[][]{$\pi/4$}
 \psfrag{C}[][]{$\pi/2$}
 \psfrag{D}[][]{$3\pi/4$}
 \psfrag{E}[][]{$\pi$}
 \psfrag{GG}[][]{$\frac{\pi}4$}
 \psfrag{GH}[][]{$\frac{\pi}2$}
 \psfrag{HG}[][]{$\frac{3\pi}4$}
 \psfrag{HH}[][]{$\pi$}
 \psfrag{C1}[][]{$c_1$}
\psfrag{C2}[][]{$c_2$}
\includegraphics[width=0.3\hsize]{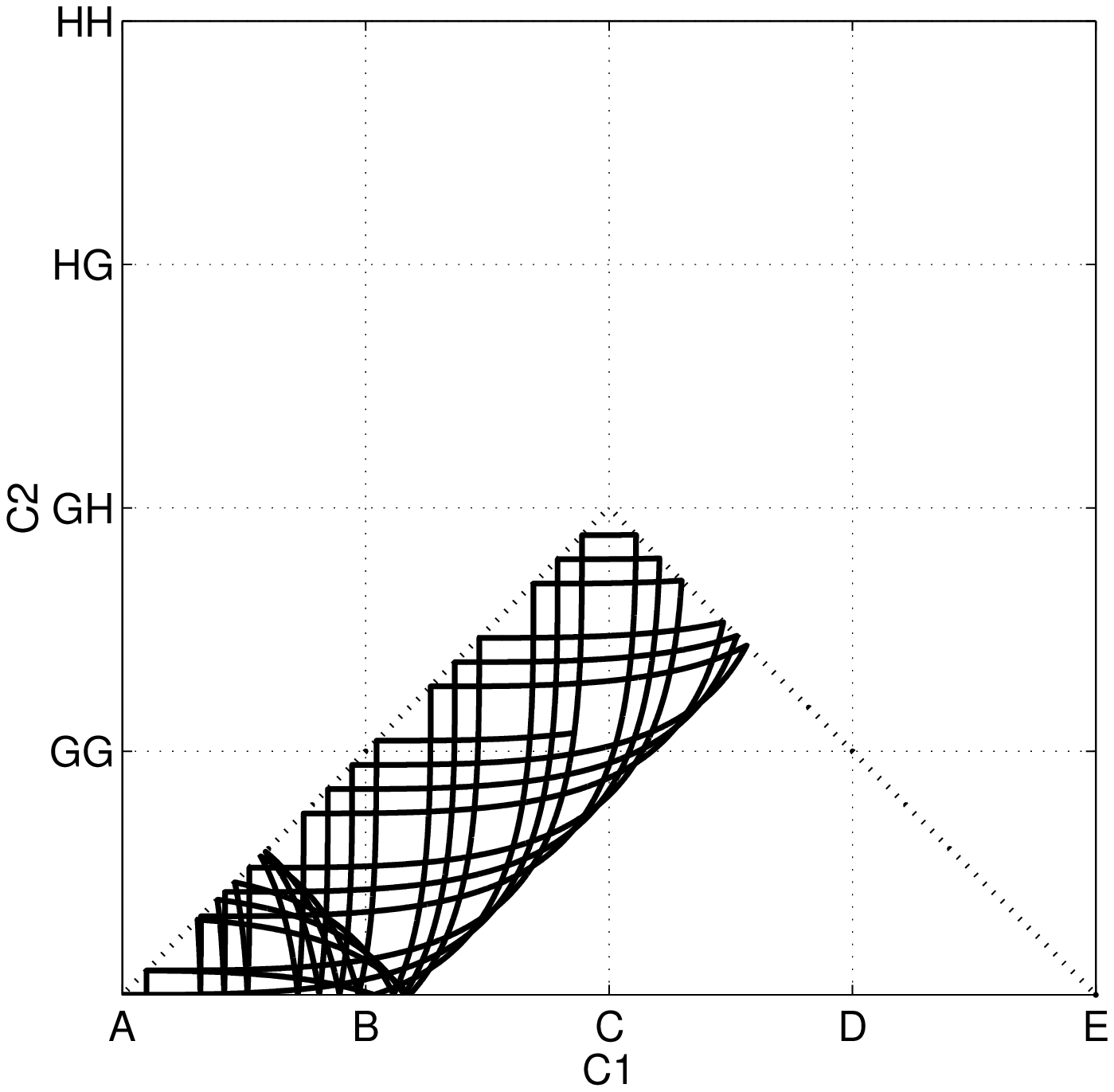}\\
$\alpha=0.5$, $t=20$ & $\alpha=0.5$, $t=80$
\end{tabular}
\end{center}
 \caption{The flow generated by $H_J$ in the Weyl chamber $\a^+$.}
 \label{fig:4}
\end{figure}
\end{example}

Note that when the Hamiltonian $H$ contains only the non-local part, that is, 
  $iH\in \p$, the flow generated by the Hamiltonian on the 3-Torus 
  has constant velocity. However, when the Hamiltonian $H$ contains
  both the local and non-local part, the velocity of the flow on the
  3-Torus is usually time dependent as shown in the above example.

\subsection{Design of universal quantum circuits}
Now let us consider how to generate {\it any} arbitrary two-qubit operation
from a given two-body Hamiltonian together with local gates. 
The local gates form the Lie subgroup $SU(2)\otimes SU(2)$, which
also contains all the single qubit operations.
We will show that by
applying the Hamiltonian at most three times, together with four
appropriate local gates, we can exactly simulate any arbitrary 
two-qubit gate, {\it i.e.}, we can implement any $SU(4)$ operation.
Consequently, if accessibility of any local gate is assumed,
this results in satisfying the universality condition needed for 
quantum computation or simulation in a very
efficient manner.

From the discussion in the preceding subsection, we know that a
given Hamiltonian is not able to generate any arbitrary two-qubit
operation simply by turning it on for a certain time period.
Generally, the set of the gates that can be generated by a Hamiltonian
directly is a one-dimensional subset of the 3-Torus.
For example, we know that $[\sqrt{\text{SWAP}}]$ can be directly
 generated from the isotropic exchange interaction between two 
physical qubits, whereas [CNOT] cannot be obtained in this way 
\cite{Makhlin:00} (unless encoding into multiple qubits is
employed \cite{DiVincenzo:00a}). 
[CNOT] can however be achieved by a circuit consisting of 
two $\sqrt{\text{SWAP}}$ and a local gate \cite{Burkard:99b}. 
We shall adopt the approach of constructing a quantum 
circuit that contain
both non-local gates generated by a given Hamiltonian and local gates,
and show that this quantum circuit can
simulate any arbitrary non-local two-qubit operation exactly with
only a small number of operations.

Consider a Hamiltonian $H$ with $iH\in \p$. The gate generated by this
Hamiltonian for a time duration $t$ is $U(t)=\exp iHt$. Consider the
following prototype quantum circuit

\begin{figure}[tb]
\begin{center}
 \footnotesize
  \psfrag{A1}[][]{$A_1$}
 \psfrag{A2}[][]{$A_2$}
 \psfrag{A3}[][]{$A_3$}
 \psfrag{O}[][]{$O$}
\psfrag{P}[][]{$P$}
 \psfrag{Z}[][]{$[\sqrt{\text{SWAP}}]$}
 \psfrag{L}[][]{$L\text{ [CNOT]}$}
\includegraphics[width=0.4\hsize]{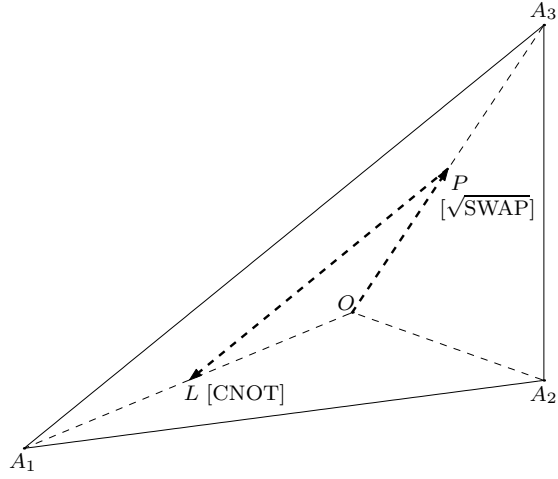} 
\end{center}
 \caption{The flow generated by the quantum circuit $k_x\exp (iH_1 t_2) k_x^\dag
  \exp (iH_1 t_1)$ in the Weyl chamber $\a^+$.}
 \label{fig:flow}
\end{figure}

\vspace{0.5cm}
\setlength{\unitlength}{0.1cm}
\begin{center}
\begin{picture}(150,10)
\put(0,0){\begin{picture}(6,10)
\put(0,2){\line(1,0){6}}
\put(0,8){\line(1,0){6}}
\end{picture}}
\put(6,0){\framebox(10,10)[c]{$k_0$}}
\put(16,0){\begin{picture}(6,10)
\put(0,2){\line(1,0){6}}
\put(0,8){\line(1,0){6}}
\end{picture}}
\put(22,0){\framebox(10,10)[c]{$U(t_1)$}}
\put(32,0){\begin{picture}(6,10)
\put(0,2){\line(1,0){6}}
\put(0,8){\line(1,0){6}}
\end{picture}}
\put(38,0){\framebox(10,10)[c]{$k_1$}}
\put(48,0){\begin{picture}(6,10)
\put(0,2){\line(1,0){6}}
\put(0,8){\line(1,0){6}}
\end{picture}}
\put(54,0){\framebox(10,10)[c]{$U(t_2)$}}
\put(64,0){\begin{picture}(6,10)
\put(0,2){\line(1,0){6}}
\put(0,8){\line(1,0){6}}
\end{picture}}
\put(70,0){\framebox(10,10)[c]{$k_2$}}
\put(80,0){\begin{picture}(6,10)
\put(0,2){\line(1,0){6}}
\put(0,8){\line(1,0){6}}
\end{picture}}
\put(86,0){\makebox(10,10)[c]{$\dots$}}
\put(96,0){\begin{picture}(6,10)
\put(0,2){\line(1,0){6}}
\put(0,8){\line(1,0){6}}
\end{picture}}
\put(102,0){\framebox(10,10)[c]{$k_{n-1}$}}
\put(112,0){\begin{picture}(6,10)
\put(0,2){\line(1,0){6}}
\put(0,8){\line(1,0){6}}
\end{picture}}
\put(118,0){\framebox(10,10)[c]{$U(t_n)$}}
\put(128,0){\begin{picture}(6,10)
\put(0,2){\line(1,0){6}}
\put(0,8){\line(1,0){6}}
\end{picture}}
\put(134,0){\framebox(10,10)[c]{$k_n$}}
\put(144,0){\begin{picture}(6,10)
\put(0,2){\line(1,0){6}}
\put(0,8){\line(1,0){6}}
\end{picture}}
\end{picture}
\end{center}
\vspace{0.5cm}
where $k_j$ are local gates, and $n$ a given integer. Note that the circuit is
to be read from left to right. The matrix representation of the above
quantum circuit is
\begin{equation}
  \label{eq:46}
k_nU(t_n)k_{n-1}\cdots k_2U(t_2)k_1U(t_1)k_0.
\end{equation}
 We will investigate the non-local gates
that can be simulated by this quantum circuit.
Recall that for this Hamiltonian $H$, 
there exists a local gate $k\in SU(2)\otimes SU(2)$ such that $\Ad_{k}
(iH)=iH_a$, where $H_a=\frac{1}2(c_1\sigma_x^1\sigma_x^2
+c_2\sigma_y^1\sigma_y^2+c_3\sigma_z^1\sigma_z^2)$.
Hence
\begin{equation}
  \label{eq:71}
   U(t)=\exp iHt= \exp(\Ad_{k^\dag}iH_a t).
\end{equation}
Let $l_j=(kk_{j-1}\cdots k_0)^\dag$, the quantum circuit 
(\ref{eq:46}) can now be described as
\vspace{0.5cm}
\setlength{\unitlength}{0.1cm}
\begin{center}
\begin{picture}(142,12)
\put(0,0){\begin{picture}(6,12)
\put(0,2){\line(1,0){6}}
\put(0,10){\line(1,0){6}}
\end{picture}}
\put(6,0){\framebox(28,12)[c]{$\exp\big(\Ad_{l_{1}} iH_a t_1\big)$}}
\put(34,0){\begin{picture}(6,12)
\put(0,2){\line(1,0){6}}
\put(0,10){\line(1,0){6}}
\end{picture}}
\put(40,0){\framebox(28,12)[c]{$\exp\big(\Ad_{l_2} iH_a t_2\big)$}}
\put(68,0){\begin{picture}(6,12)
\put(0,2){\line(1,0){6}}
\put(0,10){\line(1,0){6}}
\end{picture}}
\put(74,0){\makebox(10,12)[c]{$\dots$}}
\put(84,0){\begin{picture}(6,12)
\put(0,2){\line(1,0){6}}
\put(0,10){\line(1,0){6}}
\end{picture}}
\put(90,0){\framebox(28,12)[c]{$\exp\big(\Ad_{l_n} iH_a t_n\big)$}}
\put(118,0){\begin{picture}(6,12)
\put(0,2){\line(1,0){6}}
\put(0,10){\line(1,0){6}}
\end{picture}}
\put(124,0){\framebox(12,12)[c]{$k_nk^\dag l_n^\dag$}}
\put(136,0){\begin{picture}(6,12)
\put(0,2){\line(1,0){6}}
\put(0,10){\line(1,0){6}}
\end{picture}}
\end{picture}
\end{center}
\vspace{0.5cm}
and its matrix representation is 
\begin{eqnarray}
  \label{eq:48}
k_nk^\dag l_n^\dag\exp\big(\Ad_{l_{n}} iH_a t_n\big)
\cdots \exp\big(\Ad_{l_2} iH_a t_2\big)
\exp\big(\Ad_{l_1} iH_a t_1\big).
 \end{eqnarray}
We can then pick $k_j$  such that $l_j$ are in the Weyl group $W(G, K)$.
In that case, we have $\Ad_{l_j} iH_a t_j\in \a$. Since $\a$ is a
maximal Abelian subalgebra, the quantum circuit in Eq. (\ref{eq:48})
is locally equivalent to
\begin{equation}
  \label{eq:72}
   \exp\big(\Ad_{l_{n}} iH_a t_n+\Ad_{l_2} iH_a t_2+\cdots+
\Ad_{l_1} iH_a t_1\big).
\end{equation}
Proposition~\ref{prop:weyl} tells us that the Weyl group $W(G, K)$ is
generated by the reflections $s_\alpha$  given in Eq. (\ref{eq:37}).
Hence for a given $s_\alpha$, where $\alpha\in \Delta_{\p}$, 
there exists a local gate $k_\alpha$ such that for any $X\in \a$,
$\Ad_{k_\alpha}(X)=s_\alpha(X)$.
Following the procedure in Lemma 2.4, Ch. VII in~\cite{Helgason:78}, 
we obtain $k_\alpha$ as in the following table:
\begin{center}
\begin{tabular}{c|c|c}
\quad$\alpha$&\quad$s_\alpha([c_1, c_2, c_3])
$\quad&\quad$k_\alpha$\quad \\ \hline
\quad${i(c_3-c_2)}$&\quad$[c_1, c_3, c_2]$\quad
&\quad$\exp{\frac\pi{2}(\frac{i}2\sigma_x^1+\frac{i}2\sigma_x^2)}$\\
\quad${i(c_2-c_1)}$&\quad$[c_2, c_1, c_3]$\quad
&\quad$\exp{\frac\pi{2}(\frac{i}2\sigma_z^1+\frac{i}2\sigma_z^2)}$\\
\quad${i(c_1-c_3)}$&\quad$[c_3, c_2, c_1]$\quad
&\quad$\exp{\frac\pi{2}(\frac{i}2\sigma_y^1+\frac{i}2\sigma_y^2)}$\\
\quad${i(c_2+c_3)}$&\quad$[c_1, -c_3, -c_2]$\quad
&\quad$\exp{\frac\pi{2}(\frac{i}2\sigma_x^1-\frac{i}2\sigma_x^2)}$\\
\quad${i(c_1+c_2)}$&\quad$[-c_2,-c_1, c_3]$\quad
&\quad$\exp{\frac\pi{2}(\frac{i}2\sigma_z^1-\frac{i}2\sigma_z^2)}$\\
\quad${i(c_1+c_3)}$&\quad$[-c_3, c_2,-c_1]$\quad
&\quad$\exp{\frac\pi{2}(\frac{i}2\sigma_y^1-\frac{i}2\sigma_y^2)}$
\end{tabular}
\end{center}
Recall that the flow generated by $\exp{i H_a t}$ on the
3-Torus is $[c_1 t, c_2 t, c_3 t]$.
By choosing some appropriate $l_j$ from the Weyl group $W(G, K)$, we can steer
the flow generated by the Hamiltonian.
For example, if we want to
change the flow from $[c_1 t, c_2 t, c_3 t]$
into $[c_1 t, -c_3 t, -c_2 t]$, we can
simply apply the reflection $k_{i(c_2+c_3)}$:
\begin{eqnarray}
  \label{eq:51}
\aligned
\Ad_{k_{i(c_2+c_3)}}{i H_a t}&=
\exp{\frac\pi{2}(\frac{i}2\sigma_x^1-\frac{i}2\sigma_x^2)}
(i H_a t)
\exp{\frac\pi{2}(-\frac{i}2\sigma_x^1+\frac{i}2\sigma_x^2)}\\
&=\frac{i}2\big(c_1\sigma_x^1\sigma_x^2
-c_3\sigma_y^1\sigma_y^2-c_2\sigma_z^1\sigma_z^2\big)t.
\endaligned
\end{eqnarray}
The following example exemplifies this idea.
\begin{example}[Construction of CNOT from isotropic exchange Hamiltonian]
\label{example:flow}
Consider the isotropic Hamiltonian $H_1=\frac14(
\sigma_x^1\sigma_x^2+\sigma_y^1\sigma_y^2+\sigma_z^1\sigma_z^2)$. Our
goal is to simulate [CNOT] by a quantum circuit containing
local gates and two-qubit gates generated by $H_1$. As shown in 
Figure~\ref{fig:flow}, the
flow generated by $U(t)=\exp iH_1t$ in the Weyl chamber $\a^+$ is
$[\frac{t}2, \frac{t}2, \frac{t}2]$, which evolves along $\text{OA}_3$
for $t\in [0, \pi]$. 
In the Weyl chamber $\a^+$, the point $\text{L}$ ($[\frac{\pi}2, 0,
0]$) corresponds to [CNOT]. 
We then want to switch the flow from
$[\frac{t}2, \frac{t}2, \frac{t}2]$ to $[\frac{t}2, -\frac{t}2,
-\frac{t}2]$ at certain time instant so that the flow can reach the
point $\text{L}$. In order to do that, we can simply 
apply the reflections
$s_{i(c_2+c_3)}$ and $s_{i(c_3-c_2)}$ in series. The corresponding local gate is
thus
\begin{equation}
  \label{eq:50}
  k_x=k_{i(c_3-c_2)}k_{i(c_2+c_3)}
=\exp{\frac\pi{2}(\frac{i}2\sigma_x^1+\frac{i}2\sigma_x^2)}
\exp{\frac\pi{2}(\frac{i}2\sigma_x^1-\frac{i}2\sigma_x^2)}
=\exp{\frac{i\pi}2\sigma_x^1},
\end{equation}
and we have
\begin{equation}
  \label{eq:49}
  k_x \exp (iH_1 t) k_x^\dag=\exp \Ad_{k_x} iH_1 t
=\exp \frac{i}4(
\sigma_x^1\sigma_x^2-\sigma_y^1\sigma_y^2-\sigma_z^1\sigma_z^2)t.
\end{equation}
Now consider the following quantum circuit
\vspace{0.5cm}
\setlength{\unitlength}{0.1cm}
\begin{center}
\begin{picture}(88,10)
\put(0,0){\begin{picture}(6,10)
\put(0,2){\line(1,0){6}}
\put(0,8){\line(1,0){6}}
\end{picture}}
\put(6,0){\framebox(19,10)[c]{$\exp(iH_1t_1)$}}
\put(25,0){\begin{picture}(6,10)
\put(0,2){\line(1,0){6}}
\put(0,8){\line(1,0){6}}
\end{picture}}
\put(31,0){\framebox(10,10)[c]{$k_x^\dag$}}
\put(41,0){\begin{picture}(6,10)
\put(0,2){\line(1,0){6}}
\put(0,8){\line(1,0){6}}
\end{picture}}
\put(47,0){\framebox(19,10)[c]{$\exp(iH_1t_2)$}}
\put(66,0){\begin{picture}(6,10)
\put(0,2){\line(1,0){6}}
\put(0,8){\line(1,0){6}}
\end{picture}}
\put(72,0){\framebox(10,10)[c]{$k_x$}}
\put(82,0){\begin{picture}(6,10)
\put(0,2){\line(1,0){6}}
\put(0,8){\line(1,0){6}}
\end{picture}}
\end{picture}
\end{center}
\vspace{0.5cm}
The flow generated by this quantum circuit evolves along the line
  $\text{OA}_3$ for $t\in [0, t_1]$, and  then switches into 
a direction parallel to the line $\text{PL}$ in the plane
  $\text{OA}_3\text{A}_1$ for $t \ge t_1$. The matrix representation of
this quantum circuit is
 \begin{eqnarray}
   \label{eq:47}
\aligned
k_x\exp(iH_1 t_2)k_x^\dag\exp (iH_1 t_1)&=\exp \frac{i}4\big(
\sigma_x^1\sigma_x^2-\sigma_y^1\sigma_y^2-\sigma_z^1\sigma_z^2\big)t_2
\exp \frac{i}4\big(
\sigma_x^1\sigma_x^2+\sigma_y^1\sigma_y^2+\sigma_z^1\sigma_z^2\big)t_1\\
&=\exp\big( \frac{t_1+t_2}2 \frac{i}2\sigma_x^1\sigma_x^2
 +\frac{t_1-t_2}2 \frac{i}2\sigma_y^1\sigma_y^2
 +\frac{t_1-t_2}2 \frac{i}2\sigma_z^1\sigma_z^2 \big).
\endaligned
 \end{eqnarray}
Hence the terminal point of the flow is 
$[\frac{t_1+t_2}2, \frac{t_1-t_2}2, \frac{t_1-t_2}2]$.
 If we choose $t_1=\frac\pi{2}$ and 
$t_2=\frac\pi{2}$, the terminal point is none other than  $[\frac{\pi}2, 0,
0]$, and thus the quantum circuit simulates
[CNOT].  As shown in Figure~\ref{fig:flow}, the flow
  generated by this quantum circuit is $\text{OPL}$, which goes along
  the line $\text{OP}$ first, and after hitting the point $\text{P}$,
  it turns to the point $\text{L}$ along the line $\text{PL}$. 
Since $\text{P}$ is nothing but $\sqrt{\text{SWAP}}$, 
and $t_2=t_1=\frac\pi{2}$, 
we arrive at the known result
  that [CNOT] can be simulated by a circuit consisting of two
  $\sqrt{\text{SWAP}}$ and a local gate~\cite{Burkard:99b}. 
\end{example}

We now derive the following theorem which asserts that when $n=3$ the
quantum circuit (\ref{eq:46}) can simulate \emph{any} arbitrary
non-local two-qubit
gate. This theorem provides a geometric approach to 
construct a quantum circuit to simulate any arbitrary
two-qubit gate from a two body interaction Hamiltonian.

\begin{theorem}[Universal quantum circuit]
Given a Hamiltonian $H$ with $iH\in \p$, any arbitrary two-qubit
gate $U\in SU(4)$ can be simulated by the following
 quantum circuit
\vspace{0.5cm}
\setlength{\unitlength}{0.1cm}
\begin{center}
\begin{picture}(142,10)
\put(0,0){\begin{picture}(6,10)
\put(0,2){\line(1,0){6}}
\put(0,8){\line(1,0){6}}
\end{picture}}
\put(6,0){\framebox(10,10)[c]{$k_0$}}
\put(16,0){\begin{picture}(6,10)
\put(0,2){\line(1,0){6}}
\put(0,8){\line(1,0){6}}
\end{picture}}
\put(22,0){\framebox(18,10)[c]{$\exp(iHt_1)$}}
\put(40,0){\begin{picture}(6,10)
\put(0,2){\line(1,0){6}}
\put(0,8){\line(1,0){6}}
\end{picture}}
\put(46,0){\framebox(10,10)[c]{$k_1$}}
\put(56,0){\begin{picture}(6,10)
\put(0,2){\line(1,0){6}}
\put(0,8){\line(1,0){6}}
\end{picture}}
\put(62,0){\framebox(18,10)[c]{$\exp(iHt_2)$}}
\put(80,0){\begin{picture}(6,10)
\put(0,2){\line(1,0){6}}
\put(0,8){\line(1,0){6}}
\end{picture}}
\put(86,0){\framebox(10,10)[c]{$k_2$}}
\put(96,0){\begin{picture}(6,10)
\put(0,2){\line(1,0){6}}
\put(0,8){\line(1,0){6}}
\end{picture}}
\put(102,0){\framebox(18,10)[c]{$\exp(iHt_3)$}}
\put(120,0){\begin{picture}(6,10)
\put(0,2){\line(1,0){6}}
\put(0,8){\line(1,0){6}}
\end{picture}}
\put(126,0){\framebox(10,10)[c]{$k_3$}}
\put(136,0){\begin{picture}(6,10)
\put(0,2){\line(1,0){6}}
\put(0,8){\line(1,0){6}}
\end{picture}}
\end{picture}
\end{center}
\vspace{0.5cm}
where $k_j$ are local gates.
\end{theorem}
\begin{proof}
From Cartan decomposition of $\su(4)$ in
 Section~\ref{Sec:CartanSu4}, any arbitrary two-qubit gate $U\in SU(4)$
can be written in the following form:
\begin{equation}
  \label{eq:73}
 U=k_l  \exp\{\frac{i}{2}(\gamma_1\sigma_x^1\sigma_x^2 
 +\gamma_2\sigma_y^1\sigma_y^2+\gamma_3\sigma_z^1\sigma_z^2)\}k_r,
 \end{equation}
where $k_l$ and $k_r$ are local gates, and $\gamma_1$, $\gamma_2$,
$\gamma_3\in \R$.
We also know that for any given $iH\in
\p$, there exists a local gate $k$ such that $\Ad_{k}
(iH)=iH_a$,
 where $H_a=\frac{1}2(c_1\sigma_x^1\sigma_x^2
+c_2\sigma_y^1\sigma_y^2+c_3\sigma_z^1\sigma_z^2)$,
and  $c_1\ge c_2 \ge c_3 \ge 0$.
Therefore, the flow generated by $iH$ on the 3-Torus is $[c_1 t,
c_2 t, c_3 t]$.
The matrix representation of the above quantum circuit is
\begin{equation}
  \label{eq:85}
  k_3\exp(iHt_3)k_2\exp(iHt_2)k_1\exp(iHt_1) k_0.
\end{equation}
Let
\begin{eqnarray}
  \label{eq:86}
   \aligned
l_1&=(kk_0k_r^\dag)^\dag,\\
l_2&=(kk_1k_0k_r^\dag)^\dag,\\
l_3&=(kk_2k_1k_0k_r^\dag)^\dag,
\endaligned
\end{eqnarray}
the quantum circuit (\ref{eq:85}) can be written as
\begin{equation}
  \label{eq:87}
k_3k^\dag l_3^\dag\exp\big(\Ad_{l_{3}} iH_a t_3\big)
\exp\big(\Ad_{l_2} iH_a t_2\big)
\exp\big(\Ad_{l_1} iH_a t_1\big)k_r.
\end{equation}
Choose some appropriate local gates $k_0$, $k_1$, and $k_2$ such that
\begin{eqnarray}
  \label{eq:53}
  \aligned
l_1&=I,\\
l_2&=k_{i(c_3-c_2)} k_{i(c_1+c_3)},\\
l_3&=k_{i(c_2-c_1)} k_{i(c_3-c_2)}k_{i(c_1+c_2)},
\endaligned
\end{eqnarray}
and let $k_3=k_ll_3k$. It follows that the
 quantum circuit (\ref{eq:85}) is now 
\begin{eqnarray}
  \label{eq:54}
\aligned
\, &\quad k_l\exp\big(\Ad_{l_3} iH_a t_3\big)
\exp\big(\Ad_{l_2} iH_a t_2\big)
\exp\big(\Ad_{l_1} iH_a t_1\big)k_r\\
&=k_l\exp\big(\frac{i}2(c_3\sigma_x^1\sigma_x^2
-c_2\sigma_y^1\sigma_y^2-c_1\sigma_z^1\sigma_z^2)t_3\big)
\exp\big(\frac{i}2(-c_3\sigma_x^1\sigma_x^2
-c_1\sigma_y^1\sigma_y^2+c_2\sigma_z^1\sigma_z^2)t_2\big)\\
&\quad \cdot\exp\big(\frac{i}2(c_1\sigma_x^1\sigma_x^2
+c_2\sigma_y^1\sigma_y^2+c_3\sigma_z^1\sigma_z^2)t_1\big)k_r\\
&=k_l\exp\big((c_1t_1-c_3t_2+c_3t_3)\frac{i}2\sigma_x^1\sigma_x^2
+( c_2t_1-c_1t_2-c_2t_3)\frac{i}2\sigma_y^1\sigma_y^2
+(c_3t_1+c_2t_2-c_1t_3)\frac{i}2\sigma_z^1\sigma_z^2)\big)k_r.
\endaligned
\end{eqnarray}
To simulate the two-qubit gate $U$ in Eq. (\ref{eq:73}),
we only need to solve the following equation:
\begin{equation}
  \label{eq:55}
\left(  \begin{matrix}
    c_1&-c_3&c_3\\
c_2&-c_1&-c_2\\
c_3&c_2&-c_1
  \end{matrix}\right)
\left(
  \begin{matrix}
    t_1\\t_2\\t_3
  \end{matrix}
\right) =
\left(\begin{matrix}
  \gamma_1\\ \gamma_2 \\ \gamma_3
\end{matrix}\right).
\end{equation}
Since
\begin{equation}
  \label{eq:56}
  \det\left(  \begin{matrix}
c_1&-c_3&c_3\\
c_2&-c_1&-c_2\\
c_3&c_2&-c_1
  \end{matrix}\right)
=c_1(c_1^2-c_2c_3)+(c_1+c_2)c_3^2+(c_1+c_3)c_2^2>0,
\end{equation}
we can always find a solution for Eq. (\ref{eq:55}). Therefore, the
quantum circuit (\ref{eq:85}) can simulate any arbitrary two-qubit gate.
\end{proof}

From the above constructive proof, it is clear that  
together with four appropriate local gates,
we can simulate any
 arbitrary two-qubit gate by turning on a two-body interaction
 Hamiltonian for at most three times.
Also note that in the proof, the way to choose the local gates $k_0$, $k_1$,
 and $k_2$ is not unique. There are many different ways to choose the
 local gates and time parameters so as to
 construct the quantum circuit that achieves 
the same two-qubit operation. We therefore can pick the one
 that is optimal in terms of some cost index such as time.

\section{Conclusion}
In this paper we have derived a geometric approach to study the
properties of non-local two-qubit operations, starting from the 
Cartan decomposition of $\su(4)$ and making use of the Weyl group.  
We first showed that the geometric structure of non-local gates is a
3-Torus. By further reducing the symmetry, the geometric
representation of non-local gates was seen to be conveniently 
visualized as a tetrahedron.
Each point inside this tetrahedron corresponds to a different equivalent class of
non-local gates.
We then investigated the properties of those two-qubit operations that can 
generate maximal entanglement.  We provided a proof of the condition
of Makhlin for perfect entanglers~\cite{Makhlin:00}
 and then derived the
 corresponding geometric description of these gates
within the tetrahedral representation.  It was found that 
exactly half of the non-local two-qubit operations result 
in maximal entanglement, corresponding to a seven-faced polyhedron 
with volume equal to one half of the tetrahedron.
Lastly, we investigated the non-local operations that can be generated 
by a given Hamiltonian. We proved that given a two-body interaction 
Hamiltonian, it is always possible 
to explicitly construct a quantum circuit for exact simulation of any 
arbitrary non-local two-qubit gate by turning on the two-body 
interaction for at most three times, together with four local gates.  
This guarantees that a highly 
{\em efficient} simulation of non-local gates can be made with any
Hamiltonian consisting of arbitrary two-qubit interactions and
allowing control of single qubit operations.

\begin{acknowledgments}
We thank the NSF for financial support under ITR Grant No. EIA-0204641
(SS and KBW). JV and KBW's effort is also sponsored by the Defense
Advanced Research Projects Agency (DARPA) and Air Force Laboratory, Air
Force Materiel Command, USAF, under Contract number F30602-01-2-0524, and
the Office of Naval Research under Grant No. FDN  00014-01-1-0826.
\end{acknowledgments}

\bibliographystyle{apsrev}

\end{document}